\documentclass[article]{aa} % for the abstract without structuration 
\usepackage{multirow} 
\usepackage{natbib}
\usepackage{amssymb,amsmath}
\usepackage{graphicx, subfigure}
\usepackage{siunitx}
\usepackage{url}
\usepackage{threeparttable}
\usepackage[table]{xcolor}
\usepackage{makecell}
\usepackage{epstopdf}
\usepackage{txfonts}
\usepackage{ textcomp }

\usepackage{color, colortbl}
\usepackage{booktabs}
\usepackage{placeins}
\usepackage{tablefootnote}
\definecolor{Gray}{gray}{0.9}

\begin{document}
   \title{An empirical view of the extended atmosphere and inner envelope of the AGB star R~Doradus}
\subtitle{I. Physical model based on CO lines}

   \author{T. Khouri\inst{1}\thanks{{\it Send offprint requests to T. Khouri}
   \newline \email{theo.khouri@chalmers.se}}, H. Olofsson\inst{1}, W. H. T. Vlemmings\inst{1}, T. Schirmer\inst{1}, D. Tafoya\inst{1},
   M. Maercker\inst{1}, E. De Beck \inst{1},
   L.-\AA~Nyman\inst{1,2}, M. Saberi\inst{3}
}

\institute{Dep. of Space, Earth and Environment, Chalmers University of Technology, Onsala Space Observatory, 43992 Onsala, Sweden %1
{ \and European Southern Observatory (ESO), Alonso de Córdova 3107, Vitacura 763-0355, Santiago, Chile} %2
\and Rosseland Centre for Solar Physics, University of Oslo, PO Box 1029 Blindern, 0315 Oslo, Norway %3
}

  \abstract
  {The mass loss experienced on the asymptotic giant branch (AGB) at the end of the lives of low- and intermediate-mass
stars is widely accepted to rely on radiation pressure acting on newly formed dust grains. Dust formation happens in the
extended atmospheres of these stars, where the density, velocity and temperature distributions are strongly affected by
convection, stellar pulsation, and heating and cooling processes. The interaction between these processes and how that
affects dust formation and growth is complex. Hence, characterising the extended atmospheres empirically is paramount to advance
our understanding of the dust-formation and wind-driving processes.
}
  {We aim to determine the density, temperature, and velocity distributions of the gas in the extended atmosphere
  of the AGB star R~Dor.}
  {We acquired observations using ALMA towards R Dor
to study the gas through molecular line absorption and emission.
We use the $^{12}$CO~$v=0, J=2-1$, $v=1, J=2-1$ and $3-2$ and $^{13}$CO~$v=0, J=3-2$ lines. We modelled the observations using the 3D radiative transfer code LIME
  to determine the density, temperature and velocity distributions up to a distance of $\sim 4$ times the radius of the star at sub-mm wavelengths.}
  {The high angular resolution of the sub-mm maps
  allows for even the stellar photosphere to be spatially resolved.
%  We find the stellar continuum to peak at different positions on the stellar disc.
By analysing the absorption against the star, we infer that the innermost layer in the near-side hemisphere is mostly falling towards the star, while
gas in the layer above that seems to be mostly outflowing. Interestingly, the high angular resolution of the ALMA Band~7 observations reveal that the velocity
field of the gas seen against the star is not homogenous across the stellar disc.
The gas temperature and density distributions have to be very steep close to the star to fit the observed emission and absorption, but become shallower
for radii larger than $\sim 1.6$ times the stellar sub-mm radius. While the gas mass in the innermost region is hundreds of times larger than the mass
lost on average by R~Dor per pulsation cycle, the gas densities just above this region are consistent with those expected based on the mass-loss rate and expansion
velocity of the large-scale outflow. Our fits to the line profiles require the velocity distribution on the far side of the envelope to be on average mirrored with respect
to that on the near side. Using a sharp absorption feature seen in the CO~$v=0, J=2-1$ line,
we constrain the standard deviation of the stochastic velocity distribution in the large-scale outflow to be $\lesssim 0.4$~km/s.
We characterise two blobs detected in the CO~$v=0, J=2-1$ line and find densities substantially larger than those of the surrounding gas. The two blobs also
display expansion velocities which are high relative to that of the large-scale outflow. Monitoring the evolution of these blobs will lead to
a better understanding of the role of these structures in the mass-loss process of R~Dor.}
{}%
               
\titlerunning{R Dor seen by ALMA}
\authorrunning{T. Khouri et al.}

\maketitle
%
%________________________________________________________________

\section{Introduction}
At the end of their lives, low- and intermediate-mass stars reach the asymptotic giant branch (AGB) when an inert core rich
in carbon and oxygen has developed at the stellar centre and nuclear burning of helium and hydrogen has shifted to layers around the core \citep{Herwig2005}.
In response to the restructuring of the innermost regions, the stellar envelope becomes convective and expands { to hundreds of times larger than the stellar size in the main sequence}.
Pulsations also develop in this convective envelope and, together with the action of convective cells, create a dense, extended atmosphere.
As stars evolve along the AGB and become cooler, the temperature in the extended atmosphere becomes low enough for dust condensation
and growth to be efficient. The dust interacts strongly with the radiation field { and experiences radiation pressure.  A large-scale outflow} is triggered once enough dust has formed for
the radiation pressure force on a mass element to exceed the stellar gravitational pull \citep[e.g.][]{Hoefner2018}.

Despite the well-established qualitative picture and the substantial
progress made in the past two decades in understanding the dust-formation and wind-driving mechanism, the complexity
of the dust-formation process in the shock-impacted and non-uniform extended atmospheres still prevents the AGB mass-loss history of a star with given initial properties
to be derived from first principles.
It is therefore paramount to investigate the properties of the gas and dust in the dust-formation and wind-acceleration zones.
The picture is particularly complex for oxygen-rich AGB stars because oxygen-rich dust that is expected to survive close to AGB stars has insufficient
absorption cross-section per unit mass to drive the outflows \citep{Woitke2006}. { The current paradigm is that the nucleation of oxides starts the dust-formation process, although the exact composition
of these first condensates is still debated \citep{Plane2013,Gail2016,Gobrecht2022,Gobrecht2023}. The condensation of iron-free silicates
is thought to be necessary for the outflows to develop, with the grains needing to reach sizes $\gtrsim 0.1~\mu$m in order to provide the required opacity via scattering of photons \citep{Hoefner2008,Hoefner2016}.}
Theoretical models have advanced substantially and are now able to calculate dust growth in extended atmospheres
produced in three-dimensional simulations that take into account
convection in the envelopes \citep{Freytag2023}.

The advent of high-angular resolution instruments has revolutionised our empirical understanding of the region where dust forms and is accelerated around AGB stars. Scattered-light observations
have revealed that grains in the expected size range do exist around oxygen-rich AGB stars \citep{Norris2012,Khouri2016,Khouri2020,Ohnaka2016,Ohnaka2017,Adam2019}.
Observations of the molecular gas at high-angular resolution and sensitivity have also allowed the gas density, temperature and velocity distribution around the closest AGB stars to be studied
in unprecedented detail \citep{Wong2016,Khouri2016b,Khouri2018,Vlemmings2017,Ohnaka2019}.
Studies aiming to constrain the composition of the dust by determining gas-phase abundance of dust-forming elements have also provided important clues to the composition of
the first dust grains to form in oxygen-rich environments \citep{Khouri2014a,Khouri2018,Kaminski2016,Decin2017,DeBeck2017,Danilovich2020,Gottlieb2022}, but an 
{ empirical overview of the dust formation sequence for AGB stars} has yet to be achieved.
Despite the advancements, the physical parameters obtained from observations of the dust and gas have not yet reached
the level of providing direct testing of the feasibility of the accepted wind-driving mechanism for specific sources.

To empirically constrain the properties of the gas in the innermost layers of AGB envelopes,
we study the AGB star R Doradus using the Atacama large sub-millimetre array (ALMA).
The high angular resolution provided by ALMA
allow us to probe the crucial region where dust formation and growth take place.

This paper is structured in the following way. In Sections~\ref{sec:rdor} and~\ref{sec:obs}
we summarise the properties of R~Dor and present the observations employed in our study, respectively.
In Section~\ref{sec:obsRes} we discuss empirical results that do not rely on radiative transfer modelling. Then, we perform radiative transfer calculations
to improve the interpretation of the observations and obtain a quantitative description of the density, temperature and velocity distribution of the gas in the inner circumstellar environment of R~Dor.
The approach for calculating radiative transfer models and the results we obtain are the subjects of Sections~\ref{sec:modStrat} and \ref{sec:mod},
respectively. In Sections~\ref{sec:disc} and \ref{sec:summary}, we
discuss the results and give a summary of the paper, respectively.

\section{R Dor}
\label{sec:rdor}

R~Dor is a semi-regular variable which exhibits two modes of pulsation, with periods of 
362 and 175 days. The star switches between the two modes
on a time-scale of roughly 1000 days \citep{Bedding1998}.
Its initial mass has been estimated to
be between 1.0 and 1.3~M$_\odot$ \citep{Danilovich2017} and 1.3 and $1.6$~M$_\odot$ \citep{DeBeck2018}. In this work, we adopt an initial 
mass of 1.3~M$_\odot$.
The distance determined from the {\it Hipparcos}
parallax \citep[$62\pm3$ pc,][]{Perryman1997} is in good agreement with the distance derived by reanalysing the {\it Hipparcos} measurements,
59~pc \citep{Knapp2003}.
However, a more recent estimate suggests a smaller distance of 44~pc \citep{Andriantsaralaza2022}.
To be consistent with previous derivations of the mass-loss rate, we adopt a distance of 59~pc in this study. 

According to observations and models of the large-scale CO emission \citep{Ramstedt2014,VandeSande2018,DeBeck2018}, R~Dor is a standard semi-regular variable star with a relatively low mass-loss rate
($\sim 10^{-7}$~M$_\odot$/yr) and wind expansion velocity ($\sim 5.5$~km/s). However, when studied in
more detail, many of its characteristics are puzzling. For instance, the star and the inner circumstellar envelope seem to be rotating { hundreds of times} faster than expected for an AGB star \citep{Vlemmings2018}, the presence of a disc has
been suggested \citep{Homan2018} and has later been disfavoured \citep{Nhung2021}, { and a gas blob has been detected at tens of AU from the central star moving with a projected radial velocity at least a few times larger than the maximum
expansion velocity of the outflow} \citep{Vlemmings2018,Decin2018,Nhung2021}.
It is not clear, however, whether R~Dor is atypical in these characteristics or whether these are common features of similar sources.

\cite{Decin2018} conducted a spectral scan towards R~Dor using ALMA to reach an angular resolution of
$\sim 150$~mas. Their observations allowed
them to investigate the dynamics of the circumstellar envelope, revealing relatively weak high-velocity wings in several lines. The authors offered two
possibilities for these high-velocity outflow components: a fraction of the wind not following the smooth radial outflow, or a more forceful acceleration
of the outflow globally. 

Based on observations of polarised dust-scattered light, \cite{Norris2012} 
concluded that grains with sizes $\sim 0.3~\mu$m were responsible for the scattering
signal reported by them, while \cite{Khouri2016} derived the radial profile of the dust density distribution and concluded
that the {inferred gas-to-dust ratio at  5~R$_\star$, $\gtrsim$~5000,  is marginally consistent with the most extreme values found in
wind-driving models, of $\sim 4000$ \citep{Bladh2019}.}

Using the same Band~7 data employed in this paper, \cite{Vlemmings2019} studied the stellar continuum at $\sim 338$~GHz ($\sim 887~\mu$m).
They report a size of $\sim 62.2$~mas
 and a brightness temperature at this frequency of $\sim 2370$~K between 330.15 and 344.17~GHz.
 { We use the stellar radius measured at 887~$\mu$m, $R^{\star}_{887~\mu m}
 =1.835$~AU$=2.7\times10^{10}$~m,  as a reference size in our calculations.}
At an earlier epoch (45 days before), the stellar disc was measured to vary from 65.3 to 62.1~mas in diameter
at frequencies between 129.35 and 229.16~GHz. The authors were able to reproduce the observed variation in size as a
function of wavelength using power-law profiles for the density, temperature and ionisation. The H$_2$ density profile obtained
is $n_{\rm H_2} = 4 \times 10^{12} (r/r_\circ)^{-6}$~cm$^{-3}$,
where $r_\circ = 27.5$~mas.

\begin{figure*}[t]
% \vspace*{-2.0 cm}
\begin{center}
 \includegraphics[width=\textwidth]{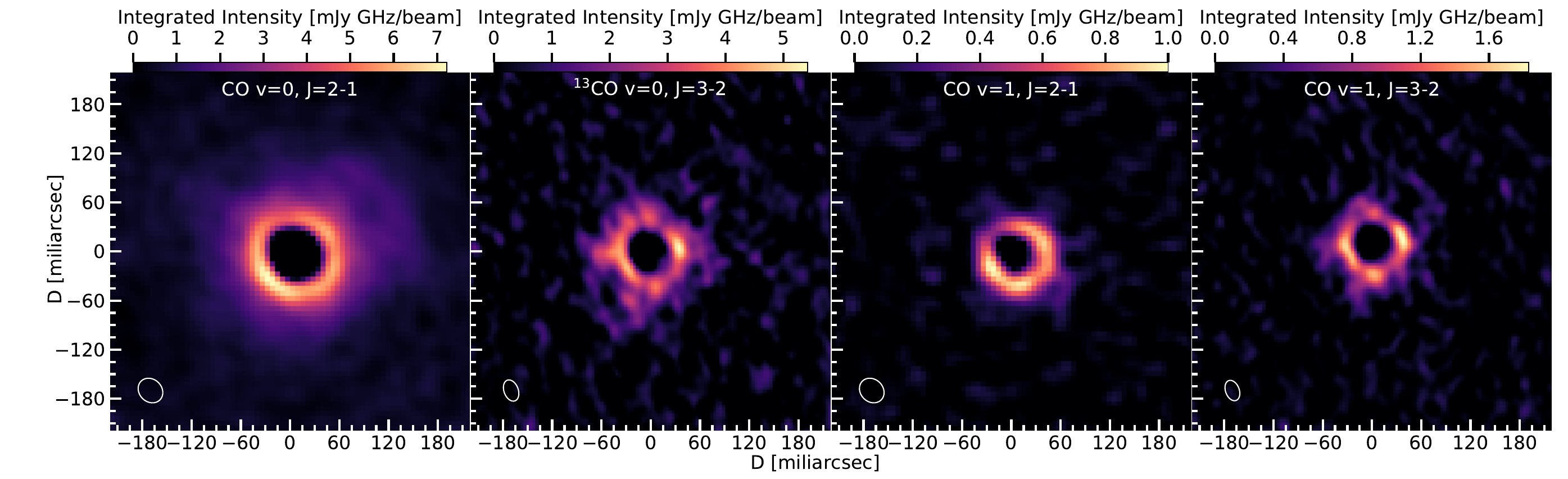} 
% \vspace*{-1.0 cm}
 \caption{Continuum-subtracted emission in the four spectral lines considered in our analysis integrated over frequency in mJy$\times$GHz/beam { (moment-zero map)}. From left to right, the lines are CO~$v=0, J=2-1$, $^{13}$CO~$v=0, J=3-2$,
 CO~$v=1, J=2-1$, and~$3-2$. The half-power sizes of the ALMA gaussian beams are indicated at the bottom left corner of each panel.}
   \label{fig:COmoment0}
\end{center}
\end{figure*}

\section{Observations}
\label{sec:obs}

R~Dor was observed using ALMA in Band 6 (project ID 2017.1.00824.S, PI Decin) on October 24
and in Band 7 (project ID 2017.1.00191.S, PI Khouri) on November 9, 2017.
The frequency range spanned by these observations covers three $^{12}$CO (hereafter referred to as only CO)
rotational transitions $v=0,J=2-1$ and
$v=1,J=2-1$ and $3-2$, and one $^{13}$CO rotational transition $v=0,J=3-2$, which we use to study the physical properties
of the gas close to R~Dor. We recovered the cubes comprising the lines mentioned above
from the products associated with the projects in the ALMA archive.
The CO~$v=1,J=3-2$ and the $^{13}$CO~$v=0,J=3-2$ lines were imaged using CASA 5.7 and the calibrated measurement sets provided by ESO.
The CO $v=0$ and $v=1$ $J=2-1$ line cubes were obtained from the ALMA archive.
The synthesised beams are $\sim 28$~mas and $\sim 20$~mas for the observations in bands 6 and 7, respectively.
The ALMA observations spatially resolve even the stellar disc of R~Dor at the two frequencies.

The ALMA observations are not sensitive to emission smooth on scales of $\gtrsim 0\farcs6$
and $\gtrsim 0\farcs42$ for the observations in Bands~6 and~7, which corresponds to regions of radius
$\sim 10$ and $\sim 7~R^{\star}_{887~\mu\rm{m}}$, respectively. However, the configuration of the array at the time of the Band~6 observations included some short
baselines, which implies that those observations can be somewhat sensitive to emission even on larger scales than the largest recoverable scale
calculated by the ALMA observatory.
The emission region of the vibrationally excited lines is expected to be
more compact than that, and, hence, the observations of these lines should not be affected by
large-scale emission being filtered out by the interferometer. This expectation is confirmed by the models we discuss below.
The lines in the ground vibrational state, however, have emission regions which are much larger than the scales probed by ALMA in the configurations
employed in the observations considered here. Hence, a very large amount of flux is expected not to be recovered.
Nonetheless, the surface brightness of the line emission decreases steeply away from the star mainly because of the gas density profile,
and the effect of resolved-out flux on the brightness distribution in the inner regions of the envelope is not necessarily strong.
Throughout the paper, we estimate the effect of resolved-out flux by simulating observations of the calculated models using the relevant
array configurations when comparing models to observations as described in Section~\ref{sec:missingFlux}.
{ As discussed in Section~\ref{sec:missingFlux},
we also use observations obtained with the Atacama Compact Array (ACA) in the context of project ID 2017.1.00595.S (PI Ramstedt)
 to constrain the large-scale emission of the low-excitation lines. The maximum recoverable scale of the maps obtained with the ACA is
 approximately $31\farcs5$. The beam has major and minor axis of $6\farcs9$ and $5\farcs0$, respectively, and a position angle of 78 degrees.}

\section{Observational results}
\label{sec:obsRes}

\subsection{Line emission}

The brightness distribution in the continuum-subtracted CO lines integrated in frequency shows a bright, thin ring
{ surrounded by weaker, more extended} emission (Fig.~\ref{fig:COmoment0}). The bright ring is the only morphological component observed in the
vibrationally excited lines and is significantly brighter than the more extended emission in the $v=0$ lines.
The brightness of the ring is fairly symmetric around the
star, with ratios between the brightest and the dimmest regions of $\sim 1.5$ for the CO $v=0, J=2-1$ and $\sim 2.5$ for the CO $v=1, J=2-1$ and $3-2$ lines.
For the $^{13}$CO line, the ratio is $\sim 2.3$.
Most of the emission is significantly closer to the average than the extremes.

The brightness distributions of the ring are somewhat different between the images
obtained at the two epochs, but roughly consistent among the lines obtained in a given epoch. The CO~$v=0, J=2-1$ and $v=1, J=2-1$ lines observed in Band~6 show a bright region to the southeast and a weaker second peak 
to the northwest (more clearly seen in the $v=1$ line). The $^{13}$CO~$v=0, J=3-2$ and CO~$v=1, J=3-2$ lines observed in Band~7 reveal peaks of emission
towards the west and east, with a third peak towards the south more clearly seen in the $v=1$ line.
The more extended emission component is very clearly seen up
to about 5~R$^{\star}_{887~\mu\rm{m}}$ in the CO~$v=0, J=2-1$ line, with a more extended bright region towards the northwest.
At 4~R$^{\star}_{887~\mu\rm{m}}$, the ratio between emission to the northwest and in other directions at the same radius varies between roughly 1.5 and 2.5.
The brightness distribution of the lines is affected by the beam
shape and by the signal to noise of the obtained images and, hence, a physical interpretation of the observed differences can be best
obtained based on models which take into account the specific different array configurations.

Observed spectra obtained in apertures of different sizes are shown in Fig.~\ref{fig:COmodel}. It is clear that in addition to line emission 
there is prominent line absorption.

\subsection{Line absorption}
\label{sec:absorption}

\begin{figure*}[t]
% \vspace*{-2.0 cm}
\begin{center}
 \includegraphics[width=\textwidth]{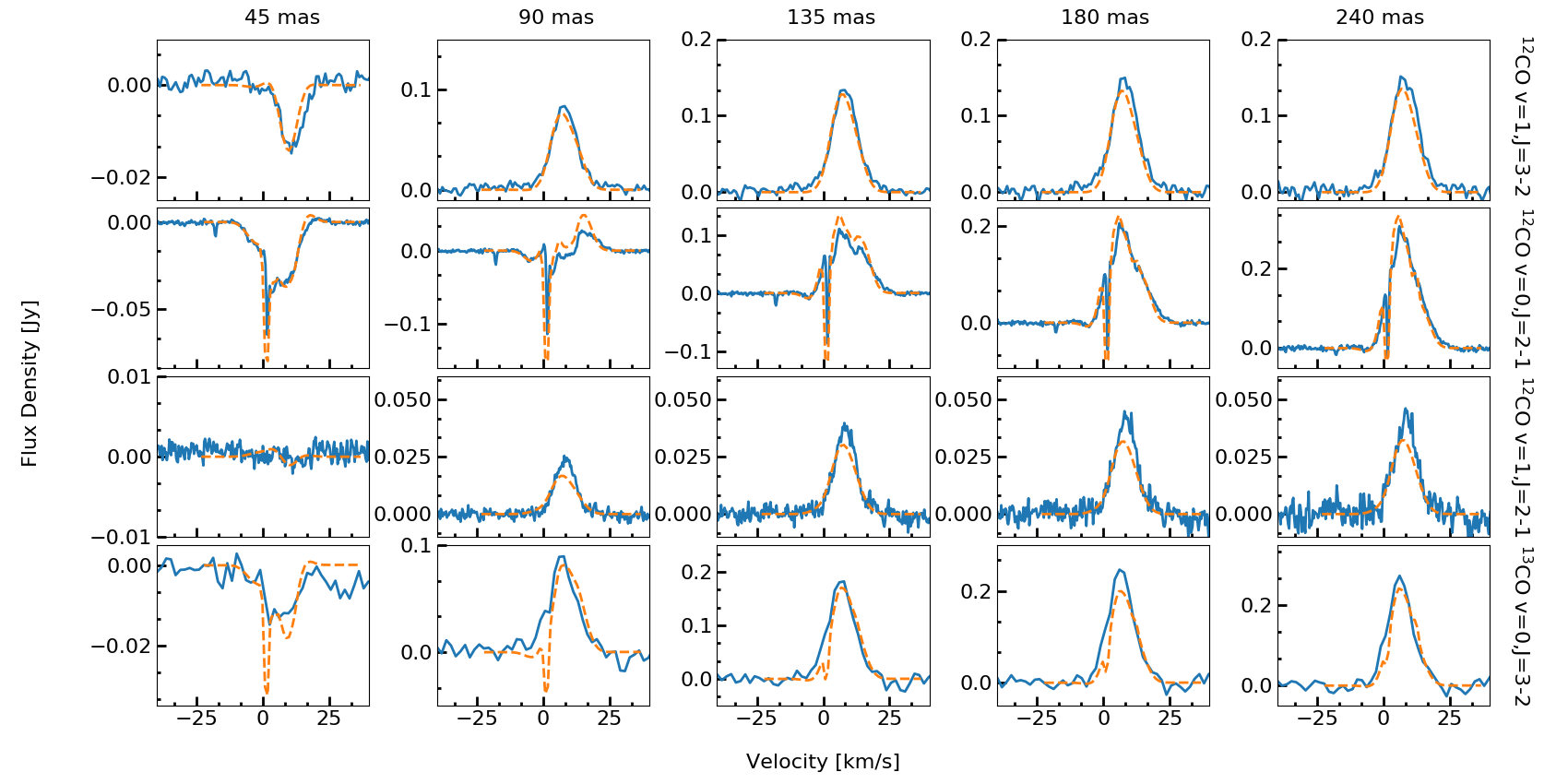}
 \caption{Extracted CO spectra. The panels show
 extracted CO spectra (blue histograms) and the best-fit model presented in Section~\ref{sec:mod}.
 The model was shifted using a systemic velocity of 7.4~km/s.
 From left to right, we present spectra extracted from apertures with diameters of 45, 90, 135, 180, and 240~mas.
 From top to bottom, the lines shown are CO~$v=1, J=3-2$, CO~$v=0, J=2-1$, CO~$v=1, J=2-1$, and $^{13}$CO~$v=0, J=3-2$.}
   \label{fig:COmodel}
\end{center}
\end{figure*}

The absorption produced by the cooler circumstellar gas against the stellar continuum is more conspicuous in
spectra extracted using apertures smaller than or comparable to the stellar radius. As the size of the aperture from which the spectrum is extracted
is increased, emission from gas in lines of sight
that do not intercept the star progressively fills up the absorption and eventually hides most of it (Fig.~\ref{fig:COmodel}).

\begin{figure}[t]
% \vspace*{-2.0 cm}
\begin{center}
 \includegraphics[width=0.5\textwidth]{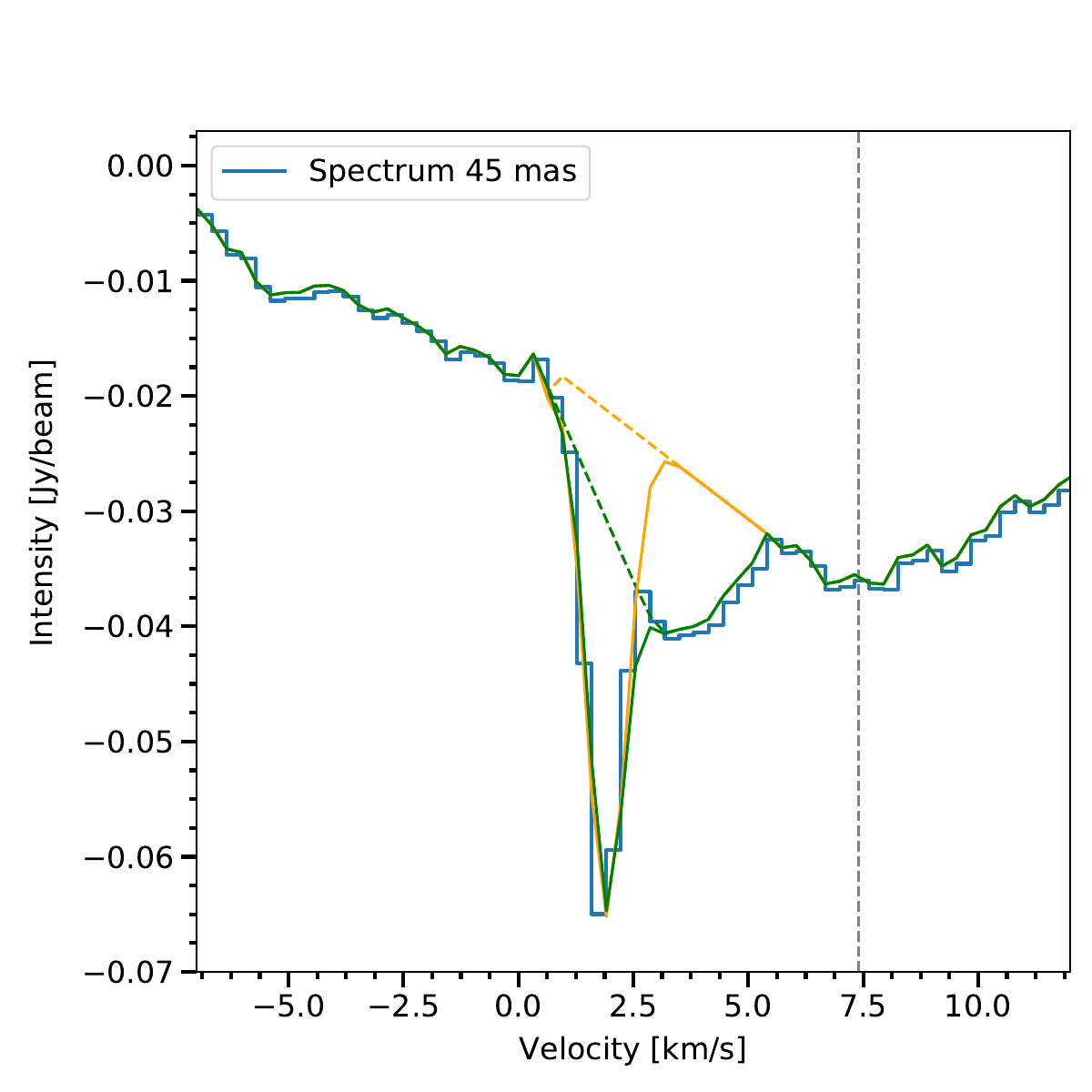} 
% \vspace*{-1.0 cm}
 \caption{Spectrum of the CO~$v=0, J=2-1$ line extracted from an aperture with a diameter of 45~mas towards the star (blue histogram). Two fits to the sharp absorption feature produced by the large-scale
 outflow are shown by the yellow and green lines. The dashed lines along the fit show our assumption for the underlying atmospheric absorption line on top of which the sharp feature is added.
 The vertical dashed line indicates the systemic velocity ($\sim 7.4$~km/s) inferred from the adopted expansion velocity of 5.5~km/s.}
   \label{fig:COturb}
\end{center}
\end{figure}

The dominant effect is from absorption by the high-density
gas close to the star, even for low-excitation lines.
Nonetheless, molecules in the outer, low-density regions of the envelope contribute to the absorption profile,
as can be seen from the very sharp absorption feature at $1.9$~km/s in the CO~$v=0, J=2-1$ line (Fig.~\ref{fig:COturb}).
This narrow feature arises from the envelope outflowing at
the terminal velocity.
Even in this case, absorption from the extended atmosphere clearly dominates the profile.

Both the absorption towards the star and the emission from the surrounding gas produce line profiles which are $\sim 30$~km/s broad (Fig. \ref{fig:COmodel}), expected to be caused by the outflow and infall of gas in the extended atmosphere
which is stirred up by convection and stellar pulsations. Such gas motions have been known to display velocities of this order from unresolved observations
of ro-vibrational transitions for many decades \citep[e.g.][]{Hinkle1978,Hinkle1982}.
Interestingly, most of the absorption observed in the CO~$v=1, J=3-2$ line
is shifted to the red (gas falling back to the star), while absorption in the CO~$v=0, J=2-1$ and $^{13}$CO~$v=0, J=3-2$ lines is more centred around the systemic velocity (Fig.~\ref{fig:COmodel}).
We interpret this as an indication that the absorbing gas is on average falling back to the star in the higher-density regions immediately above the stellar millimetre photosphere where the $v=1$ levels
are excited.

\begin{figure}[t]
   \centering
      \includegraphics[width= 9cm]{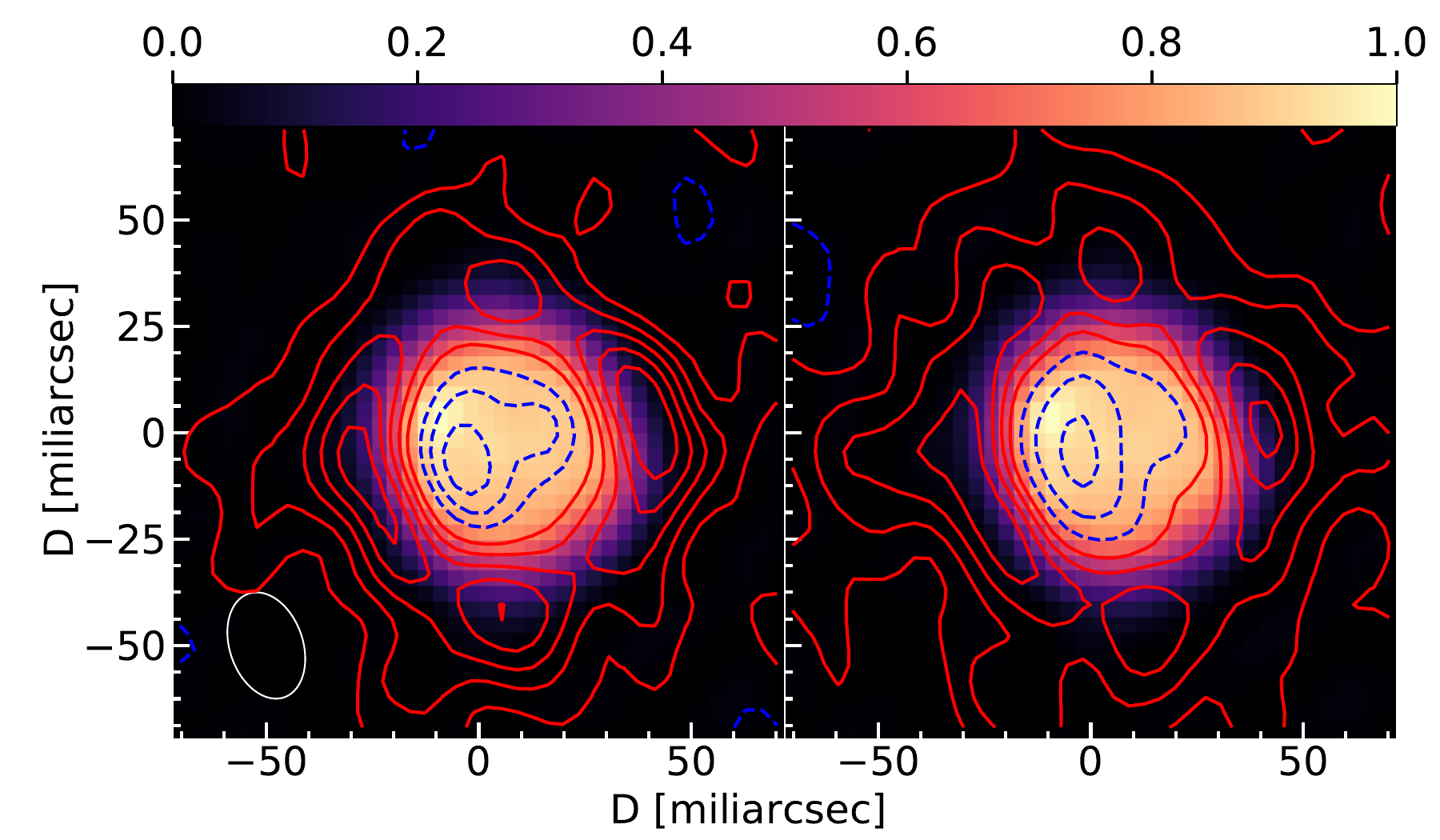}
      \includegraphics[width=9cm]{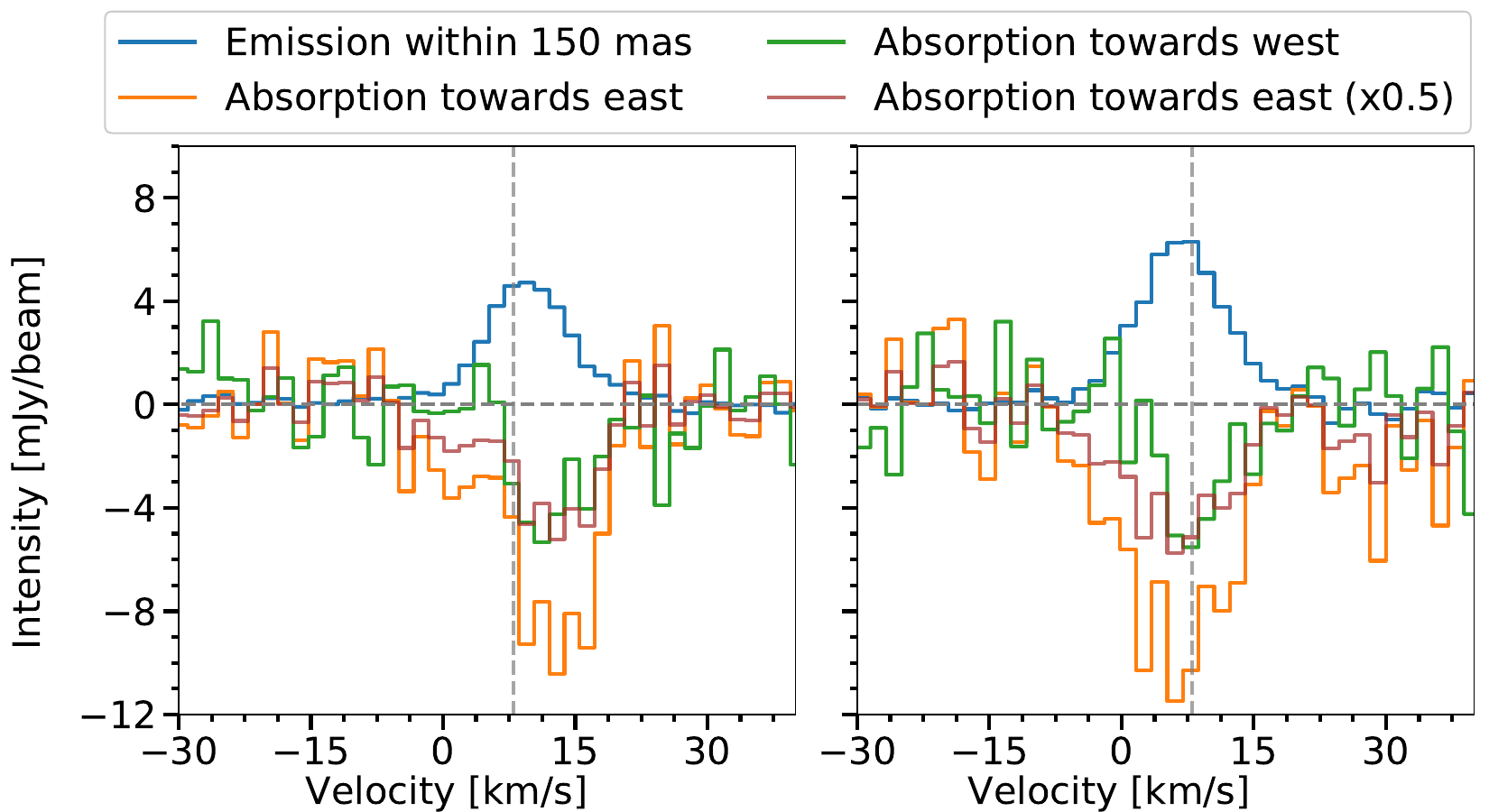}
      \caption{Upper panels: Normalised stellar continuum emission observed by ALMA in Band~7 (color maps) compared to the emission (red contours) and absorption (blue dashed contours) in the
      CO~$v=1, J=3-2$ (left panels) and $^{13}$CO~$v=0, J=3-2$ (right panels) lines.
      The positive (red) contours are drawn at 1, 2, 3, 4, and 5 mJy~$\times$~GHz/beam for both lines. The negative (blue) contours are drawn
      at -2.5, -1.5, and -0.5~mJy~$\times$~GHz/beam for the CO~$v=1, J=3-2$ line and at -3.6, -2, and -0.4~mJy~$\times$~GHz/beam for the $^{13}$CO~$v=0, J=3-2$ line.
    The root-mean-square noise is 0.4~mJy~$\times$~GHz/beam in both images.
    { Lower panels: Spectra extracted towards the absorption peak to the east of the star (yellow histogram) compared to that observed to the west (green histogram) and to the emission from an aperture
    with diameter 150~mas (blue histogram). The absorption profile to the east is also shown scaled by a factor of 0.5 in comparison (brown histogram).
    The systemic velocity determined from the fit to the absorption line obtained in Section~\ref{sec:absor_largeScale} (7.4~km/s) is shown by the vertical dahsed
    lines.}}
         \label{fig:cont_vibCO_ALMA+SPHERE}
   \end{figure}

The Band~7 observations also reveal that the { spatial distribution of the} absorption is not uniform against the stellar disc (Fig.~\ref{fig:cont_vibCO_ALMA+SPHERE}).
In these data, more absorption is seen against the eastern hemisphere
(where the 887~$\mu$m continuum emission peaks) than towards the western one.

The absorption features of  the CO~$v=1, J=3-2$ and $^{13}$CO~$v=0, J=3-2$ lines vary not only in strength between the two hemispheres but also
regarding the velocity field of the absorbing gas 
(Fig.~\ref{fig:cont_vibCO_ALMA+SPHERE}).
Towards the east, each line shows a component slightly red-shifted on average
with respect to the star that is also seen towards the west albeit about a factor of two weaker.
This component must be produced by gas falling back to the star or moving away from it at, at most,
a few km/s, and it shows more red-shifted velocities in the $v=1$ line. 
For the CO~$v=1, J=3-2$ line, a blue-shifted component seen towards the east has no counterpart above the noise level in the spectra extracted from the west.
By scaling the eastern absorption component of the CO~$v=1, J=3-2$ line to the same level as the
western one, it becomes evident that the velocity distribution of the absorbing gas is indeed different.
This blue-shifted absorption is produced by gas moving away from the star with speeds of up to $\sim 15$~km/s.
Although this difference might also be present in the $^{13}$CO~$v=0, J=3-2$ line, the noise level and the relatively bluer absorption make it
less conspicuous.

In the CO~$v=0, J=3-2$ line, we identify a weak absorption feature at $\upsilon \sim 18$~km~s$^{-1}$ which can be most easily seen in the
spectra from smaller aperture in Fig.~\ref{fig:COmodel}. To be produced by CO molecules, the absorbing gas would have to be moving at velocities of
$\sim 25$~km~s$^{-1}$ with respect to the systemic velocity. It is unclear from the data at hand whether this is indeed a feature produced
by CO molecules, or by an unidentified molecule. The detection of this absorption in other CO lines would be necessary to confirm
that this is a dynamical component of the outflow and derive the properties of the absorbing gas. Hence, we refrain from analysing this feature. 

\subsubsection{Absorption from the large-scale envelope}
\label{sec:absor_largeScale}

The narrow line profile produced by the large-scale outflow can be used to constrain the stochastic and thermal Doppler broadening in the outer envelope. In order to do this,
absorption in the inner regions of the envelope has to be disentangled from that produced in the large-scale outflow.
{ Given the unknown contribution from different components to the absorption profile shown in Fig.~\ref{fig:COturb}, an assumption on the depth of the absorption feature produced in the large-scale outflow has to be made.
Absorption by gas in the extended atmosphere dominates over most of the velocity range covered by the line profile, while gas in the wind-acceleration region might contribute significantly to the profile
at velocities between the systemic velocity and the velocity of the sharp absorption feature ($\sim 1.9$~km/s).}

{ We consider two extreme scenarios to isolate the absorption in the outer envelope:
i) the absorption feature of the outer wind corresponds to only the line that can be clearly discerned in the absorption profile (green line in Fig.~\ref{fig:COturb}), or ii)
it extends deep into the profile (yellow line in Fig.~\ref{fig:COturb}).
Our second scenario implies that absorption between the yellow and blue lines at velocities $2.5\lesssim\upsilon_{\rm LSR}\lesssim5.5$ is produced in the wind-acceleration region.
The spectra without the absorption feature from the large-scale outflow were created by interpolating the profile in the region of the feature, as shown by the dashed lines in Fig.~\ref{fig:COturb}.}

{ In order to retrieve the broadness of the absorption profile in the large-scale envelope, we fitted a gaussian function to the spectra created as described above.
The best fits were obtained using the \textsc{curve\_fit} function of the \textsc{optimize} module in the \textsc{SciPy} \citep{SciPy2020} package.
In this way, we obtain an upper limit to the stochastic velocity in the outflow.
We find standard deviations for the gaussian functions of 0.43 and 0.32~km/s for scenarios i) and ii), respectively. The uncertainties on the derived values are at most at the 2\% level.
Hence, both scenarios imply a standard deviation of at most $\sim 0.4$~km/s.}
This value is the combination of broadening from thermal and stochastic motion of the gas particles. The thermal broadening
at the expected temperatures in the outer outflow ($\lesssim 100$~K) is not necessarily negligible ($\lesssim 0.2$~km/s)
and the stochastic broadening might be significantly smaller than 0.4~km/s.
The central velocity of the fitted profile implies a
systemic velocity of 7.4~km/s given the expansion velocity of 5.5~km/s reported in the literature.

\section{Modelling strategy}
\label{sec:modStrat}

To constrain the properties and distribution of the gas, we calculated models to reproduce the observed spectral cubes using the 3D molecular excitation and
radiative transfer code LIME \citep{Brinch2010}.
We constrain the temperature, density and velocity distributions in the models by comparing
the observed CO~$v=0,~J=2-1$ and $v=1,~J=2-1$ and $3-2$, and $^{13}$CO~$v=0,~J=3-2$ lines to the calculated ones.
{ The model structure was constrained using the CO lines and the $^{13}$CO line was fit by only varying the $^{13}$CO abundance.}
We initially consider spherically symmetric models to avoid introducing a larger number of free parameters as in a more complex model.
The motivation for this approach is the fairly symmetric brightness distributions of the modelled lines, with variations in line intensity not much larger than a factor of
two for the molecular lines we consider. % { Give more details here or before?}
The only non-spherically symmetric feature we included in our models from the start is the rotation of the star and the inner circumstellar envelope reported by \cite{Vlemmings2018}.
This is implemented as a 2~km/s rotation about the south-north axis. 

The excitation of CO was calculated in non-local-thermo-dynamical (non-LTE) conditions
by considering radiative and collisional excitation in rotational levels up to $J=40$ in the ground, first and second vibrational states. The collisional rates employed
are those provided by \cite{Castro2017}. The lines predicted under non-LTE and LTE are essentially equal for the transitions in the ground vibrational state, while the
vibrationally excited lines differ. This is caused by high gas densities in these inner regions of the circumstellar envelope
and the relatively lower values of the spontaneous decay rates for the rotational transitions compared to
the ro-vibrational ones. Hence, we consider LTE for calculating the excitation of $^{13}$CO and non-LTE for calculating the excitation of CO, for which
vibrationally excited lines are also observed. Including also the $v=2$ state in our calculations affects the populations of the $v=1$ levels
in comparison to only including the $v=0$ and $v=1$ levels. Hence, we use the three vibrational states.
We adopt an abundance of CO relative to H$_2$ by number equal to $4 \times 10^{-4}$ throughout the paper.

\subsection{Missing flux}
\label{sec:missingFlux}

It is essential in our modelling to take into account the effects of observing an extended source with an interferometer.
Hence, we simulated how the { the two lines with a large emission region, CO~$v=0, J=2-1$ and $^{13}$CO~$v=0, J=3-2$,
would be seen by ALMA. We use} \textsc{CASA} version 5.7.2 and the \textsc{simalma} task to create the synthetic visibilities for
the appropriate array configurations.
These synthetic visibilities were imaged using the task \textsc{tclean} to produce the synthetic cubes.
In this way, our modelling procedure accounts for the flux missed by each of the array configurations considered.

The largest effect of resolved-out flux on the surface brightness of
small-scale structures will come from smooth, extended components with
surface brightnesses that are not insignificant compared to those of the small-scale components.
The decrease in gas densities as we move away from the star implies that the surface brightness will decrease with radius, even for
low-excitation CO and $^{13}$CO lines. Therefore, it is important that our models reproduce the observations on scales comparable to and somewhat larger than the
maximum recoverable scale.

The highest-resolution images of the CO~$v=0,J=2-1$ line towards R~Dor that can be used for this purpose,
and are available in the ALMA archive, were acquired with the
Atacama Compact Array (ACA) and the total-power telescopes on Jan 1, 2018 (project ID 2017.1.00595.S).
These ACA data were obtained about two months later than the main array observations we focus on.
The flux density recovered from the central beam { (with size $6\farcs9$ and $5\farcs0$)} peaks between 10 and 17~Jy for different velocities, which is two orders of magnitude larger
than the flux density we measure from the largest region considered by us ($0\farcs24$) of 0.15~Jy. This implies that emission from the inner region
we focus on does not affect significantly the flux density within the central ACA beam. This allows us to construct a model to reproduce the large-scale emission which is mostly
independent from the model for the innermost region. This large-scale model can, then, be used to estimate the resolved-out flux in the high-resolution observations.

\begin{table*}
\centering
\caption{Parameters considered in the models and summary of main constraints and assumptions. Positive values
of { the exponents of the temperature and density profiles, $\epsilon$ and $\eta$, respectively,}
imply temperature and density decreasing with radius. The velocities between parentheses indicate the
modified velocity field in the far-side hemisphere of the envelope.}
\label{tab:parameters}
%\begin{center}
\begin{tabular}{@{}c@{ } l l l l l l }
 Params. & \multicolumn{2}{c}{Inner} & \multicolumn{2}{c}{Intermediate} & \multicolumn{2}{c}{Outer} \\ 
 & Assumptions & Best & Assumptions & Best & Assumptions & Best \\
 \hline
$R_{\rm in}$ & $R^\star_{887\mu{\rm m}}$ & $R^\star_{887\mu{\rm m}}$  & None & $1.6 \times R^\star_{887\mu{\rm m}}$	& None \phantom{$10~R^\star_{887\mu{\rm m}}$ and 2.0 on aaaa}	& $5.0 \times R^\star_{887\mu{\rm m}}$ \\
\rowcolor{Gray}
$T_\circ$ & $T^\star$ & 	$T^\star$		 		 				& $T$ at $R^{\rm Inner}_{\rm out}$ & 940~K	& $T$ at $R^{\rm Interm.}_{\rm out}$ 			& 400~K \\
$n_\circ$ & None & $3.5\times10^{11}$~cm$^{-3}$	 		& $< n^{\rm Inner}_\circ$ & $1 \times 10^{9}$~cm$^{-3}$	& $< n^{\rm Interm.}_\circ$ 						& $4 \times 10^{7}$~cm$^{-3}$ \\
\rowcolor{Gray}
$\upsilon_\circ$ & Infall ($v=1$) &	-6.0~km/s	(10.0)				& Outflow ($v=0$) & 		13.0~km/s	 (-6.0)		& 2.0~km/s 								& 2.0~km/s \\
$\upsilon_{\rm f}$ & None & 		0.0~km/s	 (13.0)					& None & 				2.0~km/s (-2.0)			& \multicolumn{2}{l}{5.5~km/s from 10~$R^\star_{887\mu{\rm m}}$}  \\
\rowcolor{Gray}
$\epsilon$ & $> 0$ & 			2.0						 	& $> 0$ &			0.73					& $> 0$ 									& 0.6 \\
$\eta$ & $> 0$ & 				7.0							& $> 0$ & 		2.8					& \multicolumn{2}{l}{Linear acceleration up to 10~$R^\star_{887\mu{\rm m}}$} \\
%$\beta$ & 1.0 & 1.0 & 1.0 \\
\end{tabular}
%\end{center}
\end{table*}

\subsection{Physical structure of the model}
\label{sec:phys}

The steep decline in brightness of the CO lines (see Fig.~\ref{fig:COmoment0}) in the innermost regions requires models with steep density profiles.
We identify two main components in the observed maps. The first is a narrow shell around the star, seen as a ring in the images,
which accounts for emission in the vibrationally excited lines and
corresponds to the brightest region in the maps of the $v=0$ lines.
The second is only clearly visible in the maps of the CO~$v=0,J=2-1$ line and corresponds to weaker emission surrounding this inner shell.
Based on this, we included three distinct radial zones in our models. An inner one that corresponds to the inner thin shell,
an intermediate one that corresponds to the larger shell, and an outer one that corresponds to the large-scale wind. 
{ The main constraints to the parameters of density profile in all regions was} the brightness distribution of the CO~$v=0,~J=2-1$ line and
the fact that the $v=1$ lines are only observed in a thin
ring around the star.

The velocity of the gas where the different lines are produced
can be constrained using their absorption profiles observed towards the
star (see Fig.~\ref{fig:COmodel}).
Since the $v=1$ lines show only red-shifted
absorption, the velocity of the inner region was set to, on average, a slight infall towards the star. The absorption produced by this inner layer
is also clearly visible in the CO~$v=0,J=2-1$ and $^{13}$CO~$v=0,J=3-2$ lines.
To fit the observed profile of the $v=0$ lines, the velocity of the intermediate layer was set to, on average, an outflow away from the star.
This is motivated by the blue-shifted absorption which is visible in the CO~$v=0,J=2-1$ line but absent in the $v=1$ lines.

We consider a stochastic velocity in the inner and intermediate regions of 4~km/s. This is to account both for stochastic (small-scale velocity field)
and variations of gas velocity on large scales, which cannot be taken into account in a different way in a spherically symmetric model.
The temperature profile was kept continuous between the three layers, while discontinuities
were allowed in the velocity and density profiles.
Within each of the three layers, the gas temperature $T(r)$, density $n(r)$, and radial velocity $\upsilon(r)$ as a function of radius, $r$,
were approximated by the following analytical expressions

\begin{equation}
T(r) = T_\circ~\left( \frac{r}{R_{\rm in}} \right)^{-\epsilon},
\end{equation}
\begin{equation}
n(r) = n_\circ~\left( \frac{r}{R_{\rm in}} \right)^{-\eta},~{\rm and}
\end{equation}
\begin{equation}
\upsilon(r) = \upsilon_\circ+(\upsilon_{\rm f}-\upsilon_\circ) \times \left(\frac{r-R_{\rm in}}{R_{\rm out}-R_{\rm in}}\right)^{1.0},
\end{equation}
where $T_\circ$, $n_\circ$, and $\upsilon_\circ$ are the gas kinetic temperature, particle density, and radial velocity
at the inner boundary of the given layer,
$R_{\rm in}$, and $\upsilon_{\rm f}$ is the radial velocity at the outer boundary of the layer, $R_{\rm out}$.
An overview of the free parameters of the model is given in Table~\ref{tab:parameters}.

In our model for the outermost region, the gas density and radial velocity match those expected from extrapolating
inwards the mass-loss rate ($\sim 10^{-7}~{\rm M_\odot yr^{-1}}$) and expansion velocity
($\sim 5.5$~km/s) derived from models for the large-scale envelope \citep{VandeSande2018}. { The temperature profile was initially assumed
to be that given by \citep{VandeSande2018} but was modified to make it consistent with the results from the inner region,
as described below.}
We assume an expansion velocity of 2.0~km/s at
the inner radius of the outermost region of the model ($R^{\rm Outer}_{\rm in}$)
and a linear acceleration profile that reaches the maximum expansion velocity at 10~$R^\star_{887\mu{\rm m}}$.
We varied the exponent of the temperature profile in the outflow, $\epsilon_{\rm Outer}$, to fit the emission in the CO~$J=2-1$ line in the central beam
of the ACA observations.

\subsection{Modelling approach}

Our approach can be summarised as:
\begin{itemize}
\item We obtain a spherically symmetric model that fits the large
scales probed by the ACA observations.
When comparing the large-scale model and the observations, we convolved the model with the ACA beam.
This model is presented in Section~\ref{sec:ACA}.
\item We modify the density, temperature and velocity distributions
in the inner five stellar radii ($0\farcs155$ in radius) to reproduce the higher-resolution observations.
As shown in Fig.~\ref{fig:COmodel} and presented in Section~\ref{sec:firstModel},
the best model is chosen by comparison by eye to the observed line profiles extracted from five different aperture sizes (45, 90, 135, 180, and 240~mas).
\item { We adjust the large-scale model to make the temperature profile continuous, while still fitting the ACA observations. This consists of
setting $T_\circ^{\rm Outer}$ to the temperature value at the outer radius of the intermediate region. Then, $\epsilon^{\rm Outer}$ had to be modified
to fit the lines observed by the ACA.}
\item We investigate mismatches between our best model and observations that imply departure from spherical symmetry
and we modify the velocity field to address two of these issues, as discussed in Section~\ref{sec:departure}.
\end{itemize}

{ After obtaining a good fit to the data, we estimate the sensitivity of the model to variations of the input 
parameters to provide an estimate of the uncertainty in the derived physical structure.
To this goal, we present} a grid of models in Section~\ref{sec:grid} covering the 
parameter space around the parameters of the temperature and density profiles of the best models with modified velocity field.
The velocity field is not modified in the grid calculations.

As shown in Fig.~\ref{fig:COmodel} for the best-fit model, the models were compared to the data cubes for each of the four transitions considered 
by matching spectra extracted from the observed and modelled cubes using
five different aperture sizes (45, 90, 135, 180, and 240~mas), which correspond to sizes ranging from smaller than the stellar photosphere
at mm wavelengths up to $\sim 4$~stellar radii in radius.

\subsection{Central star and radiation field}

We approximate the central star by a black-body source with a temperature of 2370~K, based on the measurement of the stellar-continuum
brightness temperature from the Band~7 observations \citep{Vlemmings2019}.
This temperature has an important effect on the absorption lines seen towards the star,
because the amount of absorption depends directly on the difference between the brightness temperature of the stellar continuum and the
excitation temperature of the molecules in question.

The stellar radiation field is also important for the excitation of CO through ro-vibrational lines $\Delta v =1$ or 2
through absorption of near-IR photons with wavelengths $\sim4.6$ and $\sim2.3~\mu$m, respectively.
Therefore, the stellar infrared radiation field at these wavelengths
can have a direct effect on the level populations of the CO~$v=1$ levels relevant for our study. 
\cite{VandeSande2018} studied the spectral energy distribution of R~Dor employing an effective temperature of 2500~K, which is very close
to the value we use. We do not expect differences of a few hundred Kelvin in effective temperature to affect the stellar radiation field at
2.3 and 4.6~$\mu$m significantly, while the effect on the strength of absorption will likely be stronger. Hence, considering a black-body star
with temperature equal to 2370~K is the most appropriate for our study.

\section{Modelling results}
\label{sec:mod}

\subsection{Model for the large-scale outflow}
\label{sec:ACA}

The profile of the CO~$v=0, J=2-1$ produced by the
model chosen as representative of the brightness distribution in the region of the central beam of the ACA observations
is shown in comparison to the line observed by the ACA in Fig.~\ref{fig:CO2-1_ACA}.
Particular care was taken to select a model which produced enough emission at the systemic velocity, because the emission
region at this projected velocity
is expected to be the largest. Hence,
the problem of emission filtered out by the interferometer can be expected to be more important at this velocity.

To fit the line extracted from the central beam of the ACA observations, we find that our model requires a systemic velocity of 6.7~km/s and
a value of $\epsilon^{\rm Outer} \approx 0.6$ for a gas temperature at the inner radius of the outer region, $T^{\rm Outer}_\circ$, $\approx 400$~K,
assuming a terminal velocity of 5.5~km/s.
The value for the temperature at $R^{\rm Outer}_{\rm in}$ (5~$R^\star_{887\mu{\rm m}}$) was selected based on an initial exploration of the
gas temperatures required in the inner region to fit the lines produced in the extended atmosphere. Then, we made adjustments as required as the model for the inner
regions of the circumstellar envelope converged.

\begin{figure}[t]
% \vspace*{-2.0 cm}
\begin{center}
 \includegraphics[width=0.4\textwidth]{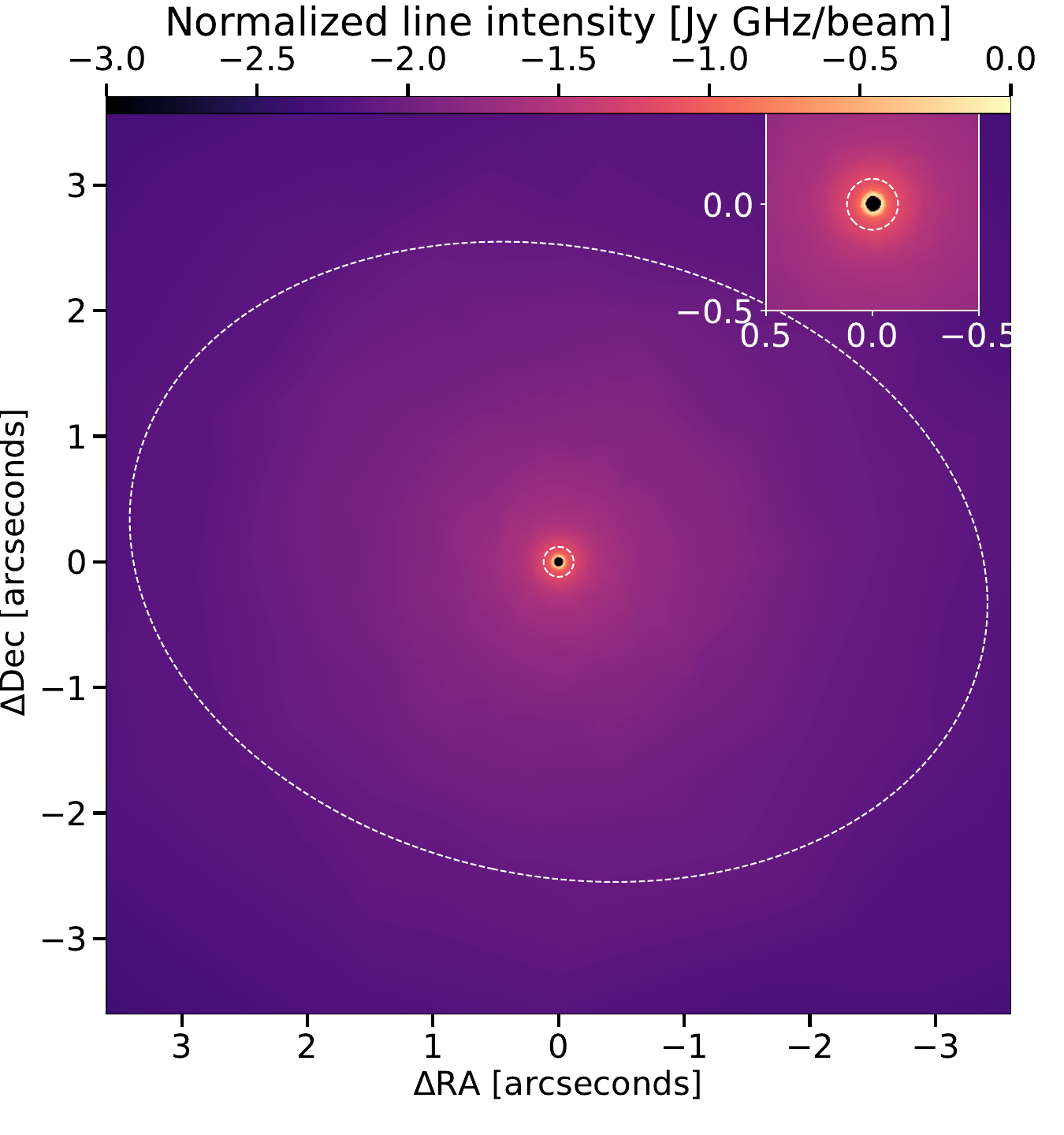} 
% \vspace*{-1.0 cm}
 \caption{Brightness distribution of the frequency-integrated CO~$v=0, J=2-1$ line in our model. The colormap shows the base-10 logarithm of the
 intensity (per pixel) normalised to the peak value in the image. In the pixels towards the star, where negative values lead to undefined logarithms, we arbitrarily set
 very negative values. The inset shows a zoom towards the inner region. { The dashed
 contours mark the largest aperture from where we extract spectra for our inner-envelope model (inner contour) and the size of the ACA beam (outer contour).
 The brightness decreases by about a factor of ten between these two contours}.}
   \label{fig:CO2-1_ACA_notConv}
\end{center}
\end{figure}

Fig.~\ref{fig:CO2-1_ACA_notConv} shows an image of the model spanning a
region comparable to the ACA beam in the observation of the CO~$v=0, J=2-1$ line. As can be seen, the surface brightness integrated over velocity decreases
by about a factor of ten between a radius of $0\farcs12$ (the maximum radius from which we extract spectra to compare to our models of the extended
atmosphere) and $3\arcsec$ (the approximate size of the ACA beam).

\subsection{Model for the inner circumstellar envelope}
\label{sec:firstModel}

The velocity, density, and kinetic temperature profiles of our best spherically symmetric
model (with rotation) for the innermost regions are shown in Fig.~\ref{fig:model}.
The departures from symmetry in the velocity field discussed in Section~\ref{sec:departure} are also indicated in the figure by the red lines.
{ We explored freely the parameter space defined by the three-zone structure presented in
Section~\ref{sec:phys} to produce this model.
The constraints discussed in Section~\ref{sec:obsRes}
allowed us to quickly narrow down the region of the parameter space that
provided good fits to the data. We found clear boundaries on the parameter space beyond which
models were unsuitable to fit the observations.
For instance, models with densities $\gtrsim 10^{10}$~cm$^{-3}$ in the intermediate region produce
emission in the $v=1$ lines from that region, which is not observed,
while models with densities $< 10^{10}$~cm$^{-3}$ in the inner region failed to
produce the observed emission in these same lines. The temperature profile in the inner region was
also well constrained by the relative amount of absorption and emission in the different lines
arising from that region.}

We find the interface between the inner and intermediate region to be located at $1.6~R^\star_{887\mu{\rm m}}$ and
that between the intermediate and outer region to be located at $5.0~R^\star_{887\mu{\rm m}}$. { However, the position of
this second interface is not well constrained, and its purpose is only to mark the transition in
the velocity profile as discussed in Section~\ref{sec:int_out}.} 
The broad lines that are produced in the extended atmosphere
(spanning more than 20~km/s) require gas distributed over a comparable velocity range ($\gtrsim 15$~km/s) in the models,
even if we use a relatively high stochastic velocity for the inner region (4~km/s).

To fit the line profiles of the inner regions, we apply a shift of 7.4~km/s to the modelled lines. This is in agreement with the value derived based on the absorption feature discussed in Section~\ref{sec:absorption}
but not with the value of 6.7~km/s derived from fitting the large-scale emission observed by the ACA discussed in Section~\ref{sec:ACA}.
The different values of the systemic velocities obtained by different methods is discussed in Section~\ref{sec:disc_vel}.

\subsubsection{The innermost region}

The innermost layer must be falling towards the star with relatively small velocities for the model to reproduce the observed absorption features in the observed lines.
This is quite well constrained because the $v=1$ levels are only significantly excited in this innermost layer. Hence, all the absorption seen in the CO~$v=1, J=2-1$ and $3-2$ lines is produced
by gas in this innermost region.
This layer also accounts for most of the absorption in the $v=0$ lines, because of the relatively high densities discussed below.
We find that infall velocities $\lesssim 6$~km/s are required to reproduce the observed velocity of the absorption features.
Small deviations between the observed and modelled profiles of the $v=0$ and $v=1$ lines are apparent.
This is likely not caused by variability because observations in each epoch include both one $v=0$ and one $v=1$ line.

To produce the ring seen in the maps of the $v=1$ lines, we find that the innermost layer in the models
must contain gas that is relatively warm and dense, with particle densities of $\sim 10^{11}$~cm$^{-3}$. The density profile is very steep
in the innermost region, with an exponent $\eta \sim 7$.
Even at these high densities, the CO~$v=1$ levels are sub-thermally populated.
To reproduce the observed ratio between absorption and emission, the models require a very steep gas temperature profile very close to the
star, with $\epsilon \sim 2$.

\begin{figure}[t]
% \vspace*{-2.0 cm}
\begin{center}
 \includegraphics[width=0.4\textwidth]{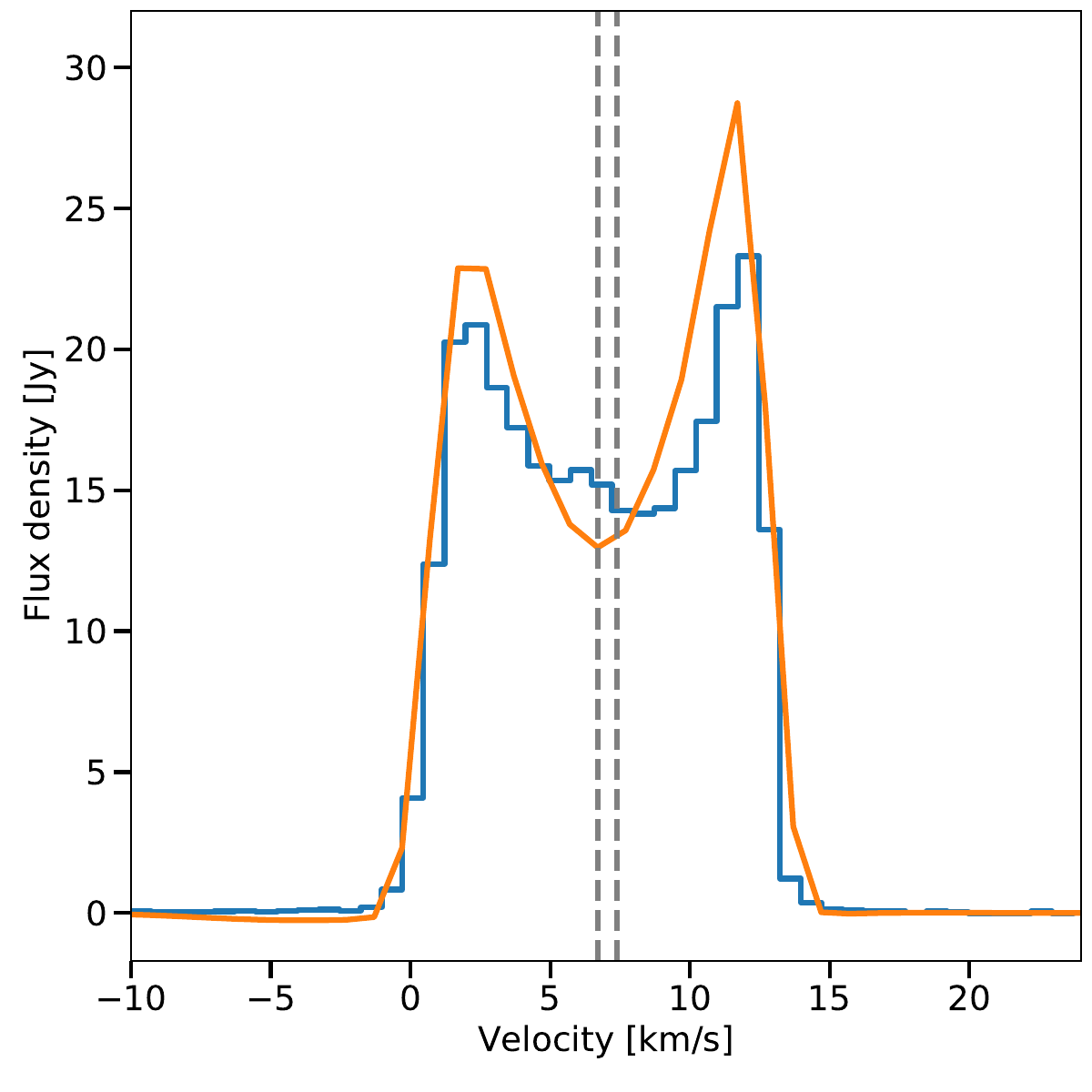} 
% \vspace*{-1.0 cm}
 \caption{CO~$v=0, J=2-1$ line extracted from the central beam of the ACA observations (blue histogram) compared to the spectrum from our best model extracted from the same region (solid orange line).
 The dashed vertical lines show the two different values derived for the systemic velocity 6.7~km/s and 7.4~km/s, which were derived, respectively,
 by fitting the position of the CO~$J=2-1$ line observed with the ACA (Section \ref{sec:ACA}) and using the central velocity of the absorption feature seen against the star (Section~\ref{sec:absorption}).}
   \label{fig:CO2-1_ACA}
\end{center}
\end{figure}

\subsubsection{The intermediate and outer regions}
\label{sec:int_out}

After the very steep radial decrease in the innermost region, the temperature profile becomes shallower, with exponents $\lesssim 0.7$ in the intermediate and outer regions.

The gas particle density decreases from values between $\sim 10^{10}$ and $\sim 10^{11}$~cm$^{-3}$ in the inner region,
to values between $\sim 10^{8}$ and $\sim 10^{9}$~cm$^{-3}$ in the intermediate region, to lower values in the outer region.
At $10~R^\star_{887\mu{\rm m}}$, the density in our model reaches values of a few times $10^{6}$~cm$^{-3}$.
The density profile becomes much shallower in the intermediate region, with $\eta \sim 3.0$  between $R^{\rm Inter.}_{\rm in}$ and $10~R^\star_{887\mu{\rm m}}$,
where we assume that the terminal velocity has been reached. In fact, both the temperature and the density profiles can be described with
a single exponent valid for both the intermediate and outer regions.
The only requirement for introducing a boundary between these two regions is the velocity profile, which we assume to increase from 2.0~km/s at  $5~R^\star_{887\mu{\rm m}}$ to
5.5~km/s at $10~R^\star_{887\mu{\rm m}}$. The relatively high expansion velocities at the intermediate layer are required to produce the blue-shifted absorption observed in the CO~$v=0, J=2-1$ line but
this higher-velocity gas could be located at { a different radial distance within the intermediate
region without affecting our model significantly}. Hence, the exact shape of the velocity profile in the intermediate and outer
regions is not very well constrained and, especially in the outer region, { depends strongly on our assumptions.}

\begin{figure*}[t]
% \vspace*{-2.0 cm}
\begin{center}
 \includegraphics[width=\textwidth]
{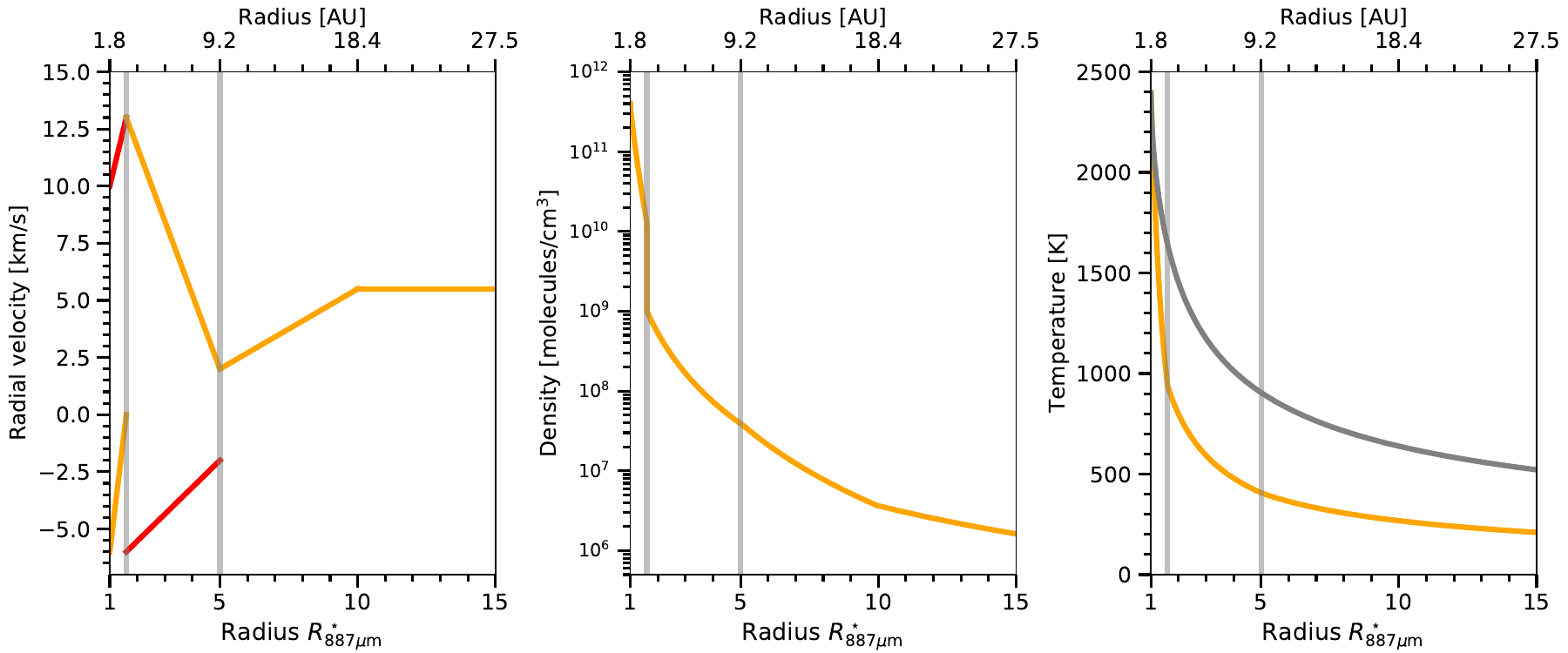}
% \vspace*{-1.0 cm}
 \caption{Physical structure of the best model shown in Fig.~\ref{fig:COmodel} and presented in Section~\ref{sec:mod}. The orange lines show the profiles of radial velocity (left panel), density (middle panel), and temperature (right panel)
 in the symmetric model with rotation and the red lines show the modifications introduced in the velocity field in the far-side hemisphere to improve the fit to the line profiles. The tangential (rotational) velocity
 is not indicated in these plots. The grey line in the temperature panel shows a grey atmosphere temperature profile.
 The vertical light grey lines mark the transition between the inner and intermediate regions and between the intermediate and outer regions in the model (as shown in Table~\ref{tab:parameters}).}
   \label{fig:model}
\end{center}
\end{figure*}

\subsection{The $^{13}$CO line}

The only parameter varied when fitting the observed $^{13}$CO~$v=0, J=3-2$ line was the $^{13}$CO abundance. We initially assumed a value of 10 for the $^{12}$CO/$^{13}$CO
ratio as reported in the literature \citep{Ramstedt2014}, which implies a $^{13}$CO abundance of $4 \times 10^{-5}$ relative to H$_2$. However, such a model produces too-strong absorption and emission compared to the observations.
The model for R~Dor presented by \cite{Ramstedt2014} consisted of a mass-loss rate of $1.6 \times 10^{-7}$~M$_\odot$/yr and a
CO abundance of $2\times 10^{-4}$, which differ by a factor of 1.6 and 2.0 from the mass-loss rate and CO abundance adopted by us.
We obtain a much better fit (shown in Fig.~\ref{fig:COmodel}) using a $^{13}$CO abundance of $2.7 \times 10^{-5}$ relative to H$_2$, which corresponds to a $^{12}$CO/$^{13}$CO of 15.
Given the uncertainties in the models and the different mass-loss rates and molecular abundances between our study and that of \cite{Ramstedt2014}, the discrepancy is not concerning.

\subsection{Effects of departures from spherical symmetry in the line profiles}
\label{sec:departure}

The observations show multiple deviations from our symmetric model with rotation: a velocity-shift between the $v=0$ 
and $v=1$ lines, mismatching contributions to the  red-shifted $^{12}$CO~$v=0$ emission, and two emission blobs. 
We adjust our model to address the former two, as we interpret these mismatches as being connected to the 
description of the velocity field. We describe the mismatches and our modelling strategy to overcome them here. We 
do not attempt to model the blobs in the outflow, but characterise them in Section~\ref{sec:blobs}.

\subsubsection{Shift between centre velocities of CO lines in the $v=0$ and $v=1$ states}
\label{sec:shiftv1}

\begin{figure}[t]
%   \centering
      \includegraphics[width= 8cm]{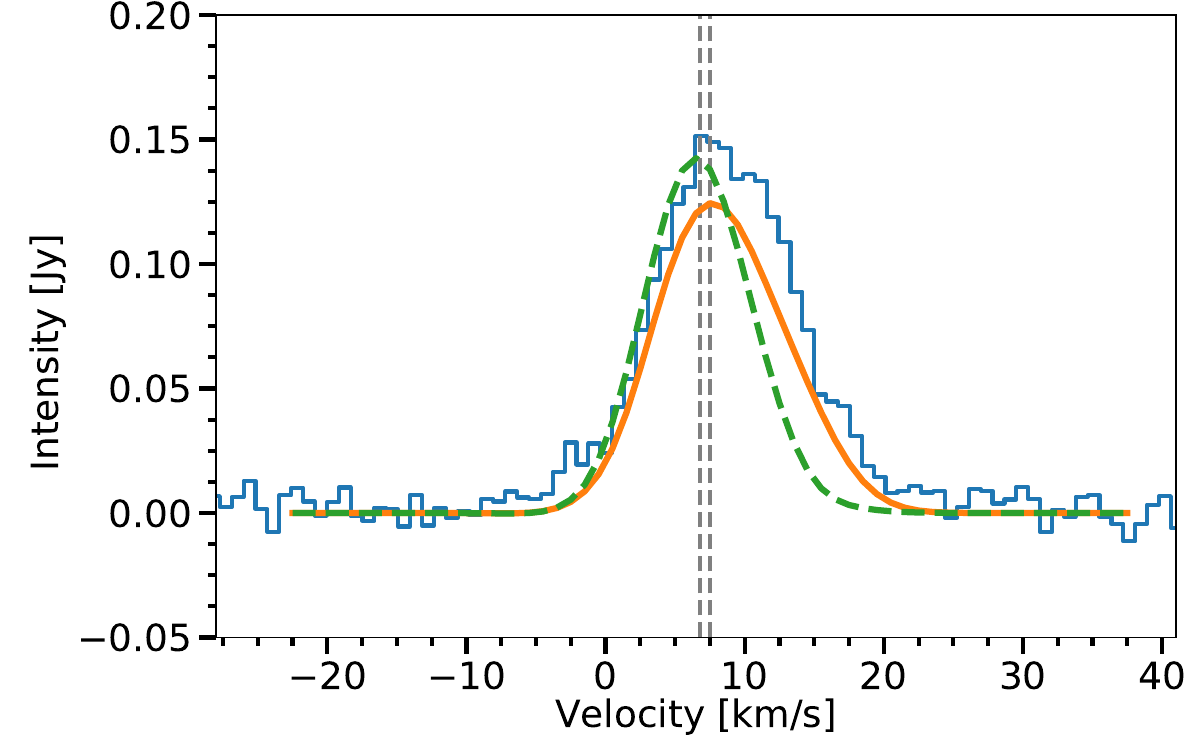}
%\vspace{-0.9cm}      
      \includegraphics[width= 8cm]{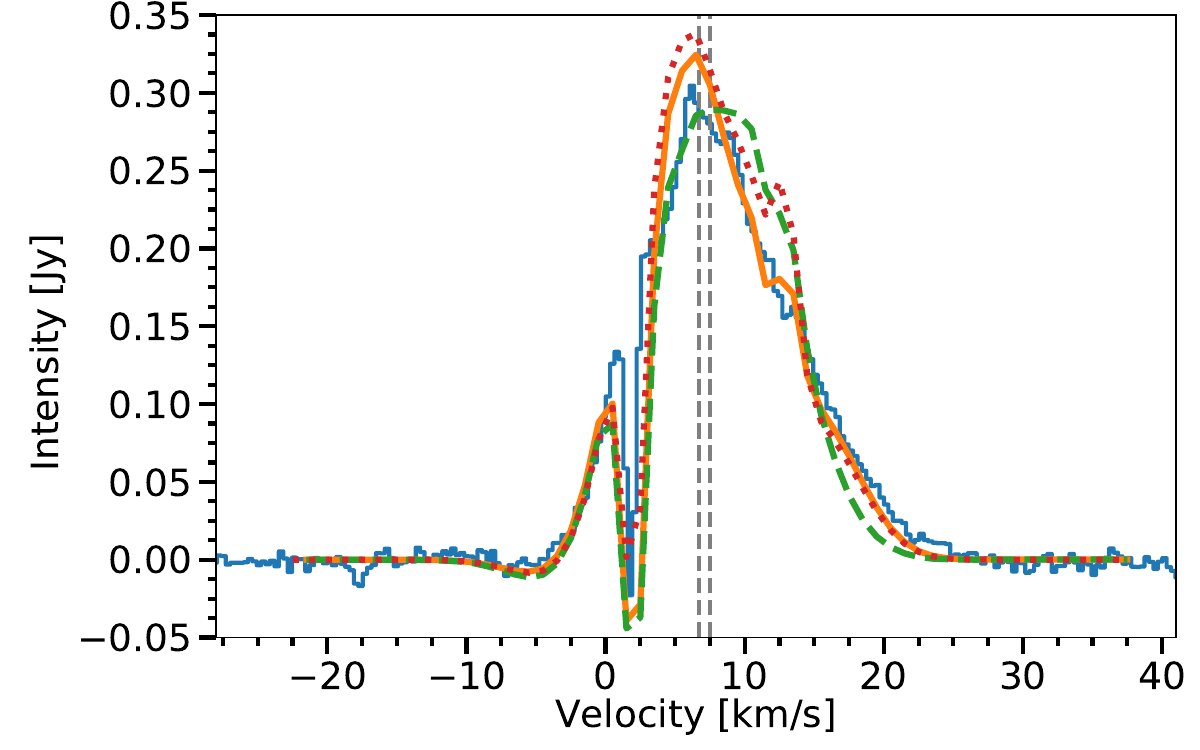}
      \caption{ Comparison between the CO~$v=1, J=3-2$ (upper panel) and CO~$v=0, J=2-1$ (lower panel) line profiles of the model with symmetric radial velocity field (green dashed line) and the model with an outflow with the modifications to the velocity
      field discussed in Sections~\ref{sec:shiftv1} and \ref{sec:shiftv0} (yellow solid line). The observed spectra are shown by the blue histograms and the possible values of the systemic velocity are indicated by the
      vertical dashed gray lines. We adopt a systemic velocity of 7.4~km/s when shifting the model to fit the data.
      All spectra were extracted from an aperture with 240~mas diameter centred on the star.
      In the lower panel, the spectrum of the
      model with the modifications to the velocity field is shown without simulating of observation by the ALMA array (dotted red line). In this case, only smoothing to the beam resolution was carried out. Hence, the comparison between the red dotted
      line and the yellow solid line shows how much flux is being resolved out by simulating the interferometric observations.}
         \label{fig:symm_vs_nonSymm}
   \end{figure}

When we compare both the emission profiles extracted from larger regions and the absorption profiles, we find a deviation between the prediction
from our spherically symmetric model with rotation and the observed
velocity of the centre of the CO~$v=0, J=2-1$ and CO~$v=1, J=3-2$ lines (Fig.~\ref{fig:symm_vs_nonSymm}). An inspection of the
CO~$v=1, J=3-2$ { channel maps} reveals a much more symmetric emission ring at velocities of $\sim 13.5$~km/s in the observations
({ Fig.~\ref{fig:COv1_channels}}) than the model with symmetric velocity distribution and rotation ({ Fig.~\ref{fig:modelCOv1_symm}}). In order to reproduce the observed brightness distribution, more
emission is required at that (red-shifted) velocity to the east of the star.

To improve the fit to both the line profile and the resemblance of the channel maps of the CO~$v=1$ lines,
we modified the velocity field of the innermost region
of the far-side hemisphere { (with respect to the Earth)}
by substituting the infall by an outflow with a velocity of $\sim 12$~km/s. This shifts the centre of the CO~$v=1$ lines by the required amount.
Interestingly, the absorption profile shown in Fig.~\ref{fig:cont_vibCO_ALMA+SPHERE}
also reveals outflow and infall in the inner layers towards the east of the star, indicating that at least parts of the inner layer are indeed moving away
from R~Dor. Hence, even if our approximation of an outflowing inner layer in the far-side of the envelope is somewhat rough, the existence
of outflowing gas in this inner layer which is mostly not seen in absorption is plausible based on the data.

\begin{table}
%\centering
\caption{Range of parameters explored in the models and steps used. In the third column, it is indicated by $+$ or $\times$ whether the successive
values of the given parameter were obtained by adding or multiplying, respectively, the previous value and the step size.}
\label{tab:grid}
%\begin{center}
\begin{tabular}{l l l l }
Parameter & Min. & Step & Number of steps \\
\hline
 \multicolumn{4}{c}{Inner Region} \\
$n_\circ^{\rm Inn}$ & $3.3\times 10^{10}$ & $\times3.3$ & 5 \\
$\eta_\circ^{\rm Inn}$ & 5 & $+1$ & 5 \\
$\epsilon_\circ^{\rm Inn}$ & 1.0 & $+0.5$ & 5 \\
\hline
\multicolumn{4}{c}{Intermediate Region} \\
$n_\circ^{\rm Inter.}$ & $1.0\times 10^{8}$ & $\times3.3$ & 5 \\
$\eta_\circ^{\rm Inter.}$ & 1 & $+1$ & 5 \\
$\epsilon_\circ^{\rm Inter.}$ & 0.37 & $+0.15$ & 5 \\
$T_\circ^{\rm Inter.}$ & 540 & $+200$ & 5\tablefootnote{An extra set of models was added with $T_\circ^{\rm Inter.} = 400$~K, increasing to a total of six grid
steps in this parameter.} \\
\end{tabular}
\end{table}

\subsubsection{CO~$v=0, J=2-1$ line profile}
\label{sec:shiftv0}

Our model also fails to reproduce the CO~$v=0, J=2-1$ emission line profiles because emission is not produced at the correct velocities on the red-shifted side of the line.
This is particularly noticeable for spectra extracted from apertures with diameters larger than about 120~mas, when emission dominates with respect to the absorption { (e.g. Fig.~\ref{fig:symm_vs_nonSymm})}.
To suppress emission at the required velocities, we modified the velocity field in the intermediate region by changing the outflow in the
far-side hemisphere to a slight infall, with velocities $\lesssim 6$~km/s.

\subsection{Model sensitivity assessment}
\label{sec:grid}

{ To assess the sensitivity of the model to the input parameters and}
quantify the uncertainties of the derived temperature and density profiles, we ran two grids of models.
In these grids, we varied the density and temperature at the inner radius of the inner and intermediate shells
and the exponents of the temperature and density profiles. Contrary to the models discussed up to this point,
we did not require the temperature profile to be continuous in these
calculations. { We do not expect this approach to produce physically viable models, but our intention
is to better isolate the effects of varying different parts of the temperature profile. If the temperature profile
is required to be continuous, changes to one region imply modifications to the others. This makes
interpreting the effects of variation in one parameter more difficult.}

The parameters of each region were varied in two independent grids because otherwise a very large number of model calculations would be 
necessary. For each of the grids, the parameters of the region not being modified were kept the same as those of the best model 
(Section~\ref{sec:firstModel}) for that region. The velocity field in all models was the same and includes the modifications introduced in 
Sections~\ref{sec:shiftv1}  and \ref{sec:shiftv0}. The parameters of the outer region were always kept the same as the best model 
(Section~\ref{sec:firstModel}).

\begin{figure*}[t!]
% \vspace*{-2.0 cm}
\begin{center}
 \includegraphics[width=0.95\textwidth]{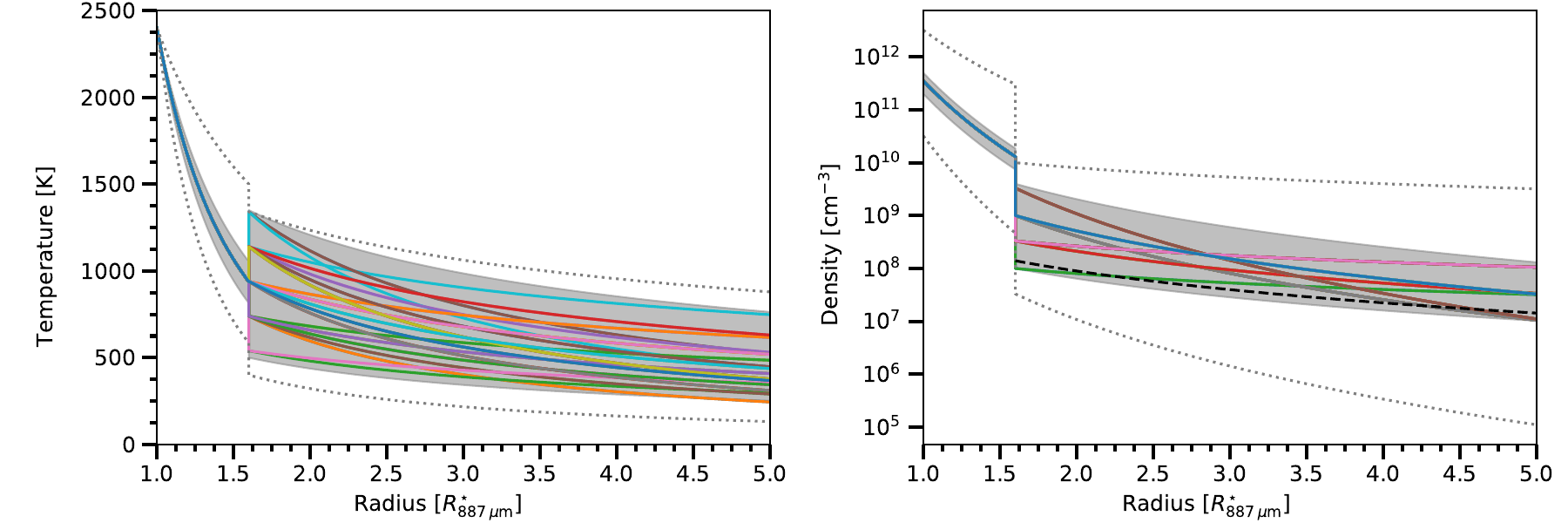} 
\end{center}
\caption{Models that provide an acceptable fit to the observed CO line profiles (solid lines). The grey region indicates the region over which the values of gas temperature and density as a function
of radius can vary in the context of our model and still fit the data. The black dashed line indicates the density of an outflow with a constant velocity of 5.5~km/s and a mass-loss rate of $10^{-7}$~M$_\odot$/yr. The dotted lines represent the limits of the range explored in the grid calculations
for the  temperature and density profiles.}
 \label{fig:grid}
\end{figure*}

{ The range of parameter space explored in the two grids of models is given in Table~\ref{tab:grid} and is shown in Fig.~\ref{fig:grid}.
The grid over the parameters for the inner and intermediate regions consisted of 125 and 750 models, respectively.
T$^{\rm Inn}_\circ$ was kept fixed at $T^\star_{\rm 887~\mu m}$ (2370~K).}

The models were ranked based on the sum of the squares of the residuals obtained from subtracting a given grid model from the best model from 
Section~\ref{sec:firstModel}. The comparison was done using the two lines which provided the strongest constraints,
CO $v=0, J=2-1$ and $v=1, J=3-2$,
extracted from the apertures with 45, 90, 135, 180, and 240~mas
in diameter (same line spectra as shown in Fig.~\ref{fig:COmodel}).
Both models were convolved with the respective ALMA beam
before subtraction, but we did not perform simulation of ALMA observations with the respective array configuration for these comparisons.
This approach was chosen because simulating the ALMA observations is time consuming and simulating all models from the grid would require a 
large amount of computation time.
 Since the brightness distribution of the extended outflow was kept unchanged
and we aim to minimise differences in the brightness distribution of the inner regions, the effect of resolving-out flux would be
very similar for any good model.
Therefore, by comparing grid models to the best model in this way,
we avoid the long computation time without introducing any biases or shortcomings in our model selection.

Finally, we compared the top-ranked models by eye to judge whether the fits were indeed acceptable. 
This approach was employed because the complexity of the model makes the ranking based on squared residuals too simplistic.
We discuss the model-selection procedure with examples in more detail in the Appendix~\ref{app:grid}.

Only one model from the grid for the inner region provided a good fit to the data, while 52 models from the grid for the intermediate region provided good fits.
Hence, the model for the innermost region is very well constrained, with all acceptable models producing the same set of parameters.
The model for the intermediate region has a larger range of acceptable values.
We attribute this to the stronger constraints { provided by emission detected both in the $v=1$ and $v=0$~CO lines in the inner region,
while only the CO~$v=0, J=2-1$ line is seen in the intermediate region.}
This single transition does not allow us to break the degeneracy between density and temperature distributions for the intermediate region.
The radial profiles of the temperature and density of the selected grid models are shown in Fig.~\ref{fig:grid}. On average, these models show lower densities in the intermediate region
than our best model from Section~\ref{sec:firstModel}.

The density of an outflow with constant velocity of 5.5~km/s and mass-loss rate of $10^{-7}$~M$_\odot$/yr is also shown in Fig.~\ref{fig:grid} by the black dashed line.
For a steady-state wind, the average density as a function of radius
in the inner regions would be expected to be higher than this line. Since the density in an outflow with constant mass-loss rate scales proportionally to the inverse of the expansion velocity, gas densities would be below or above
this line for faster or slower gas velocities, respectively. In a steady-state outwards-accelerating outflow, the dashed line corresponds to a lower limit for the gas density as a function of radius.

\begin{figure}[t]
% \vspace*{-2.0 cm}
\begin{center}
 \includegraphics[width=0.45\textwidth]{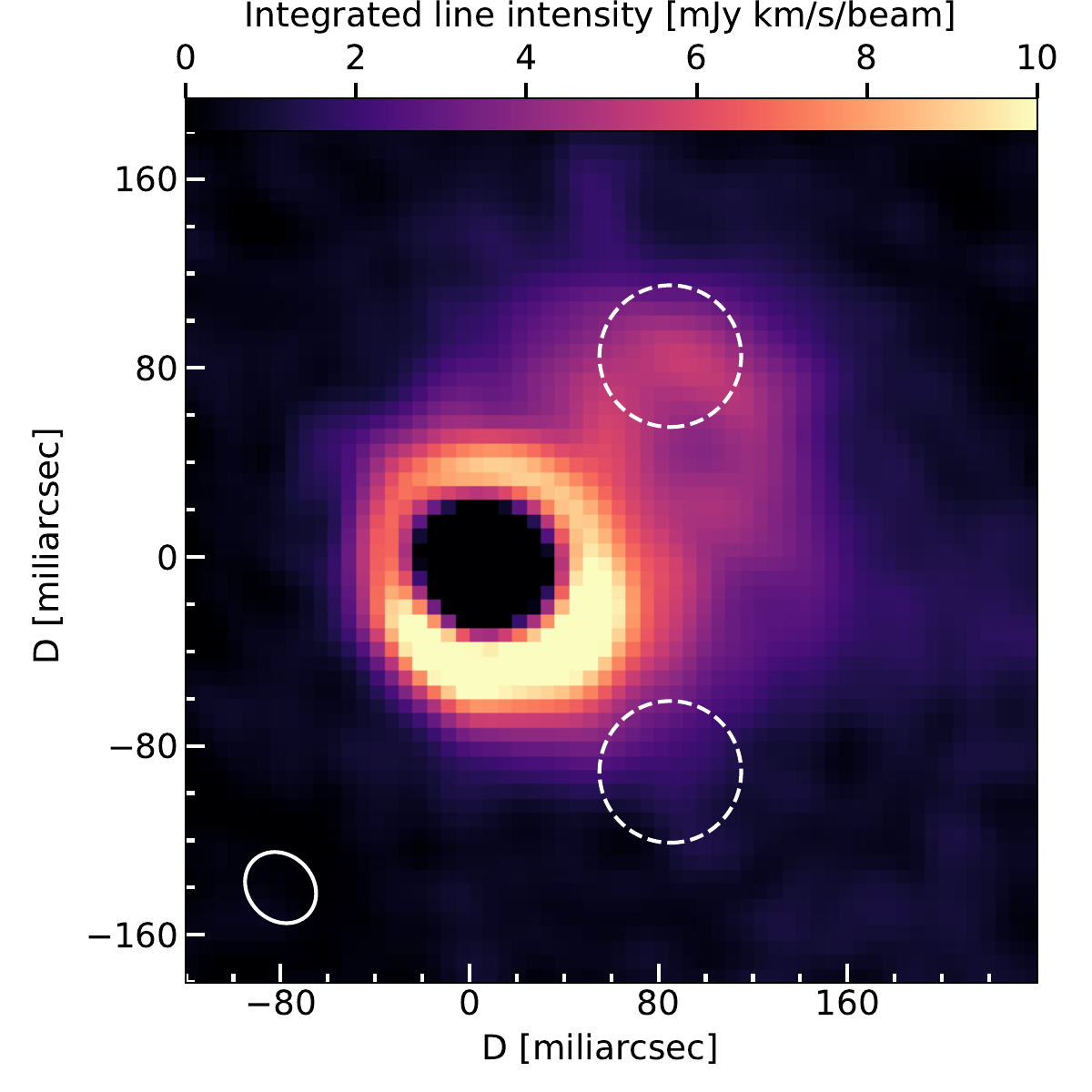} 
\end{center}
\vspace*{-0.9 cm}
\begin{center}
 \includegraphics[width=0.45\textwidth]{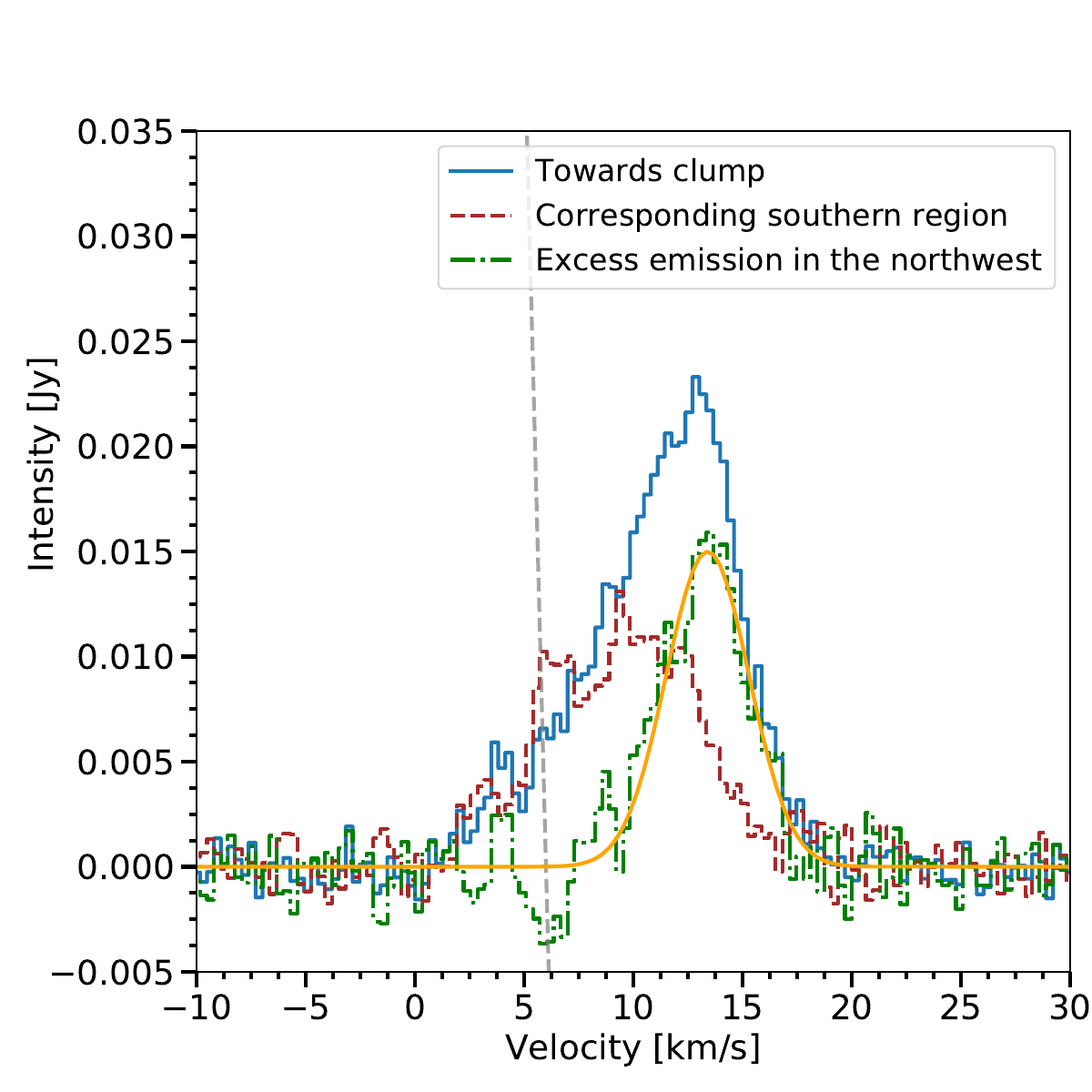} 
% \vspace*{-1.0 cm}
 \caption{Upper panel: { integrated line intensity (moment-zero)} map of CO~$J=2-1$ emission integrated in velocity between 9~km/s and 19~km/s. The dashed white circles shows the region from which we extracted spectra to characterise the blob (circle on the northern hemisphere)
 and the region for the reference spectrum (circle on the southern hemisphere).
 The white ellipse shows the half-power size of the ALMA beam.
 { Lower panel: spectrum towards the inner blob (blue solid line) and a corresponding region in the southern hemisphere (brown dashed line). The dash-dotted green line shows
 the difference between these two spectra, the solid yellow gaussian shows the fit to the excess emission in the northwest. The gaussian function
 has an amplitude of 15~mJy, a standard deviation of 1.9~km/s and is centred at 13.4~km/s.
 For reference, the
 vertical dashed gray line indicates one of the values for the systemic velocity derived by us ($7.4$~km/s).}}
   \label{fig:innerClump}
\end{center}
\end{figure}

\subsection{Blobs}
\label{sec:blobs}

We identify two blobs of gas are much brighter in the CO~$v=0, J=2-1$ line than their surroundings in
the ALMA observations. The first blob has a small projected distance and seems to be located very close to the extended atmosphere
to the northwest of the star. The second, seems to lie at a significantly larger distance from the star in the southeast direction.

\subsubsection{Inner northwest blob}

This blob blends in the images with emission from the inner and intermediate layers. It is
located roughly between $0\farcs08$ (4.7~AU, 2.6~$R^{\star}_{887~\mu\rm{m}}$) and $0\farcs15$ (8.85~AU, 4.8~$R^{\star}_{887~\mu\rm{m}}$)
from the centre of R~Dor and has an extent in the tangential direction of roughly 110~mas, as shown in Fig.~\ref{fig:innerClump}.
Although an inspection of the channel maps ({ Fig.~\ref{fig:COv0_channels}}) suggests that this blob is indeed located close or even partially inside the extended atmosphere, it is not
certain that this is not a projection effect.

Isolating the blob emission from that of the gas along the same column is
difficult because of its projected location and velocities very close to those of the high-density regions of the extended atmosphere.
Instead of attempting to integrate emission over the whole
blob, we extracted spectra towards the blob using an aperture of 60~mas in diameter with a centre placed at $0.\arcsec12$
(7.1~AU, 3.85~$R^{\star}_{887~\mu\rm{m}}$)
from the centre of the star towards the northwest. This radius and aperture size were chosen so that the whole aperture lies within the blob while
avoiding significant contamination from the bright ring of the extended atmosphere.
We also extracted the spectrum towards a corresponding aperture (same size and distance from the star) towards the southwest (away from blob).
This southwest position was chosen because the spectra would be expected to be equal to that of the region towards the blob,
given the sphericity and { the south-north} rotation axis considered in the model.
We find the observed CO~$v=0,J=2-1$ line has a flux of $(1.32\pm0.03)\times 10^{-21}$~W/m$^{2}$ and a full-width at half maximum (FWHM)
of $8.5\pm0.3$~km/s towards the blob and a flux of $(0.80\pm0.02)\times 10^{-21}$~W/m$^{2}$ and a FWHM of $8.9\pm0.3$~km/s
away from the blob. The spectra are shown in Fig.~\ref{fig:innerClump}.

We isolate the blob emission from that of the more extended outflow by subtracting the spectrum obtained towards the
region in the southwest (away from the blob) from that obtained towards the northwest (towards the blob). 
For this offset-corrected spectrum, we obtain the fairly gaussian line shown in
Fig.~\ref{fig:innerClump}, which has an amplitude of 15~mJy, a central velocity of 13.5~km/s,
a FWHM of 4.9~km/s, and an integrated flux of $\sim 5 \times 10^{-22}$~W/m$^2$.

To estimate the excitation temperature, 
we use eight SO$_2$ lines, six in Band~7 and two in Band~6, identified in the spectrum extracted from the 60~mas aperture. { We measured
the line fluxes by integrating the spectra over the observed line profiles. The obtained line fluxes were used to calculate the column density of molecules in the
upper level, $N_{\rm u}$, for each transition. We derived an excitation temperature by fitting the distribution of level populations of SO$_2$ molecules
following the procedure described in \cite{Goldsmith1999}. The best fit and the associated uncertainty were obtained using the
\textsc{curve\_fit} function of the \textsc{optimize} module in the \textsc{SciPy} package.}  We
find an excitation temperature of SO$_2$ of
$545\pm113$~K (green points in Fig.~\ref{fig:SO2}).

Based on the temperature profile in our model, this gas temperature is reached at $\sim 3.4~R^{\star}_{887~\mu\rm{m}}$ from the star.
Considering the one-$\sigma$ interval in excitation temperature, we find a range in radii between $\sim 2.6~R^{\star}_{887~\mu\rm{m}}$ and
$\sim 4.6~R^{\star}_{887~\mu\rm{m}}$.
The upper range of this estimated radial distance implies the angle between the position vector of the blob and the plane of the
sky to be at most $\sim 30~^\circ$.
Taking into account the line-of-sight velocity of the blob with respect to the systemic velocity ($\sim 7$~km/s), we find
a radial velocity of $\sim 13$~km/s and an ejection timescale $\sim 2$~years.
If the blob is closer to the plane of the sky, its radial velocity increases considerably. For 15~$^\circ$, we find a radial expansion velocity
of $\sim 25$~km/s. Hence, angles between the position vector of the blob and the plane of the sky smaller than 15~$^\circ$
imply relatively high velocities and are unlikely.

{ Assuming optically thin emission towards the blob, an excitation temperature of CO of $550$~K implies an excess gas mass of
$\sim 3 \times 10^{-8}$~M$_\odot$ is necessary to reproduce the emission
measured within the 60~mas aperture towards the northwest with respect to the one towards the southwest (green line in Fig.~\ref{fig:innerClump}).} 
Assuming this region to be representative of the density of the blob and approximating the blob by an ellipse with
minor and major axes equal to 70 and 110~mas, we find a total excess mass of $\sim6 \times 10^{-8}$~M$_\odot$. Assuming
the size of the blob along the line of sight to be 5.3~AU (the average between the axes of the ellipse considered above),
we find an excess H$_2$ density of $\sim 10^8$~cm$^{-3}$. This compares to densities of $8\times 10^{7}$~cm$^{-3}$
and $5\times 10^{7}$~cm$^{-3}$ in the 
CO model at 3.85 and 4.45~$R^{\star}_{887~\mu\rm{m}}$, respectively.
Hence, the gas density within the blob is between a factor of two to three
times that at a similar distance from the star to the southwest.

\subsubsection{High-velocity southeast blob}
\label{sec:highV_blob}

\begin{figure}[t]
% \vspace*{-2.0 cm}
\begin{center}
 \includegraphics[width=0.45\textwidth]{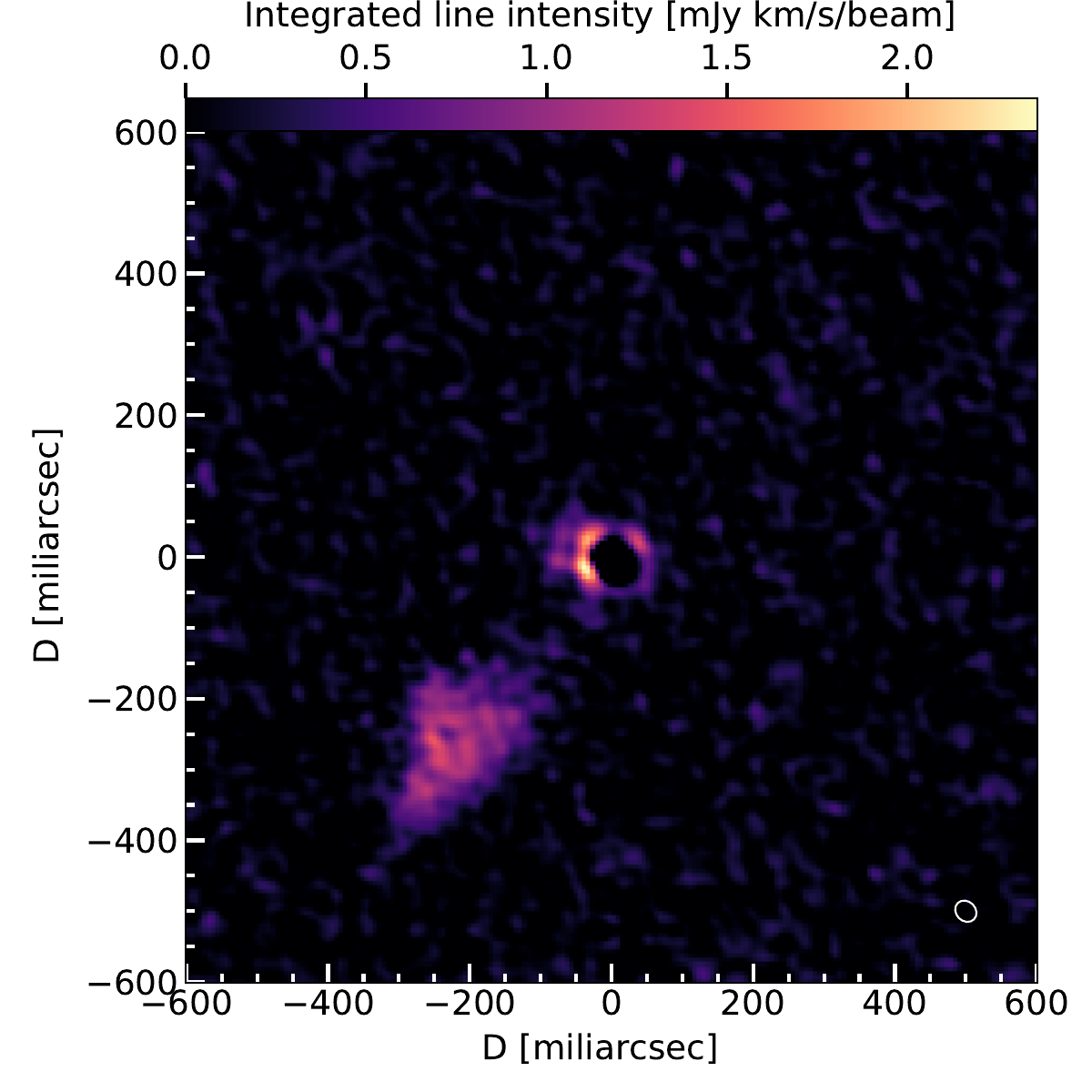} 
\end{center}
\vspace*{-0.9 cm}
\begin{center}
 \includegraphics[width=0.45\textwidth]{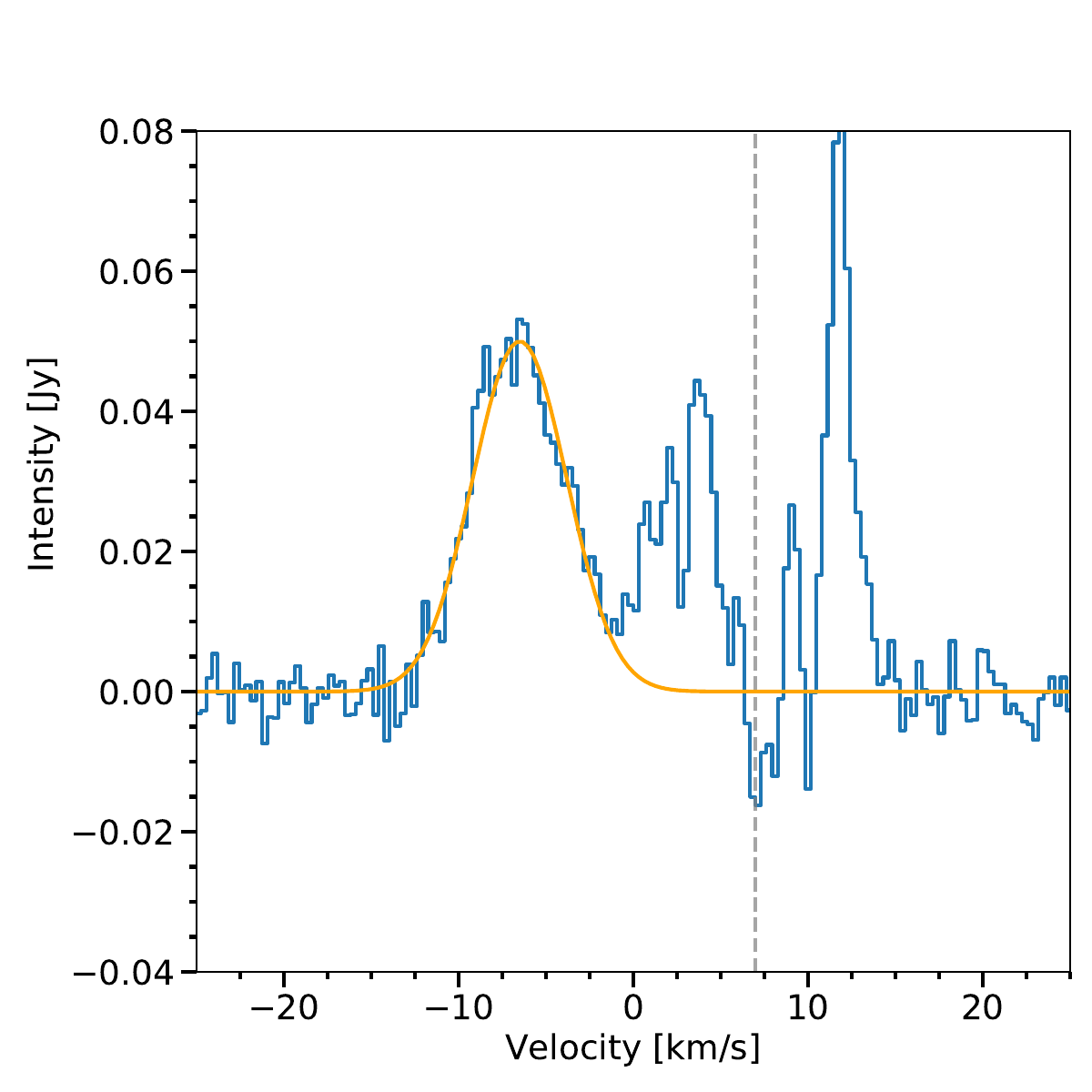} 
% \vspace*{-1.0 cm}
 \caption{Upper panel:  { integrated line intensity (moment-zero)} map of the CO~$v=0, J=2-1$ emission integrated in velocity between -14 and -1~km/s. The white ellipse
 in the lower right corner shows the size at half maximum of the ALMA
 gaussian beam. Lower panel: spectrum extracted towards the region defined by the blob in the upper panel (blue histogram). A gaussian fit to the emission associated with the blob is shown in the solid orange line. { For reference, the
 vertical dashed gray line indicates one of the values for the systemic velocity derived by us ($7.4$~km/s).}}
   \label{fig:CO_highVel}
\end{center}
\end{figure}

The blob to the southeast of R~Dor, shown in Fig.~\ref{fig:CO_highVel}, has a projected radial distance of $\sim 0\farcs36$
(21.5~AU, 11.7~$R^{\star}_{887~\mu\rm{m}}$)
from the centre of the star. Even if this blob is considerably outside of the region we consider when fitting our model (up to $0\farcs12$,
or 4~$R^{\star}_{887~\mu\rm{m}}$, in radius around the star),
its peculiar characteristics make the derivation of its properties potentially relevant for understanding the wind-driving mechanism.
%Hence, we investigate its properties in detail below.

Using the spectrum extracted from an aperture of $0\farcs14$ diameter towards the blob, we find an average line-of-sight velocity of the blob {$\sim~-14$~km/s} with respect to the systemic velocity.
The line profile has a full-width at half maximum of $\sim 6.5$~km/s, as can be seen in the spectrum extracted towards the blob (Fig.~\ref{fig:CO_highVel}).
We interpret the velocity along the line of sight to be the projection of a radial expansion velocity.
An alternative explanation is that the blob is in orbit and has a tangential, rather than radial, velocity around R~Dor. The expected orbital velocities at
21.5~AU from a 1.3~M$_\odot$ star is $\sim 7$~km/s, considerably less than the observed velocity along the line of sight. Moreover, considering the low probability of observing
a blob at an edge-on orbit near its point of maximum projected velocity along the line of sight, the scenario in which this is a gas blob orbiting R~Dor is unlikely.

A radially moving blob with projected negative velocity can be located either at the near side of the
circumstellar envelope and moving away from R~Dor (outflow), or at the far side of the envelope and falling towards R~Dor (infall). Given that
the flow velocity gets more negative with increasing distance from the star, the explanation that the blob is moving away from R~Dor (and towards us)
is favoured. 

\begin{figure}[t]
% \vspace*{-2.0 cm}
\begin{center}
 \includegraphics[width=0.45\textwidth]{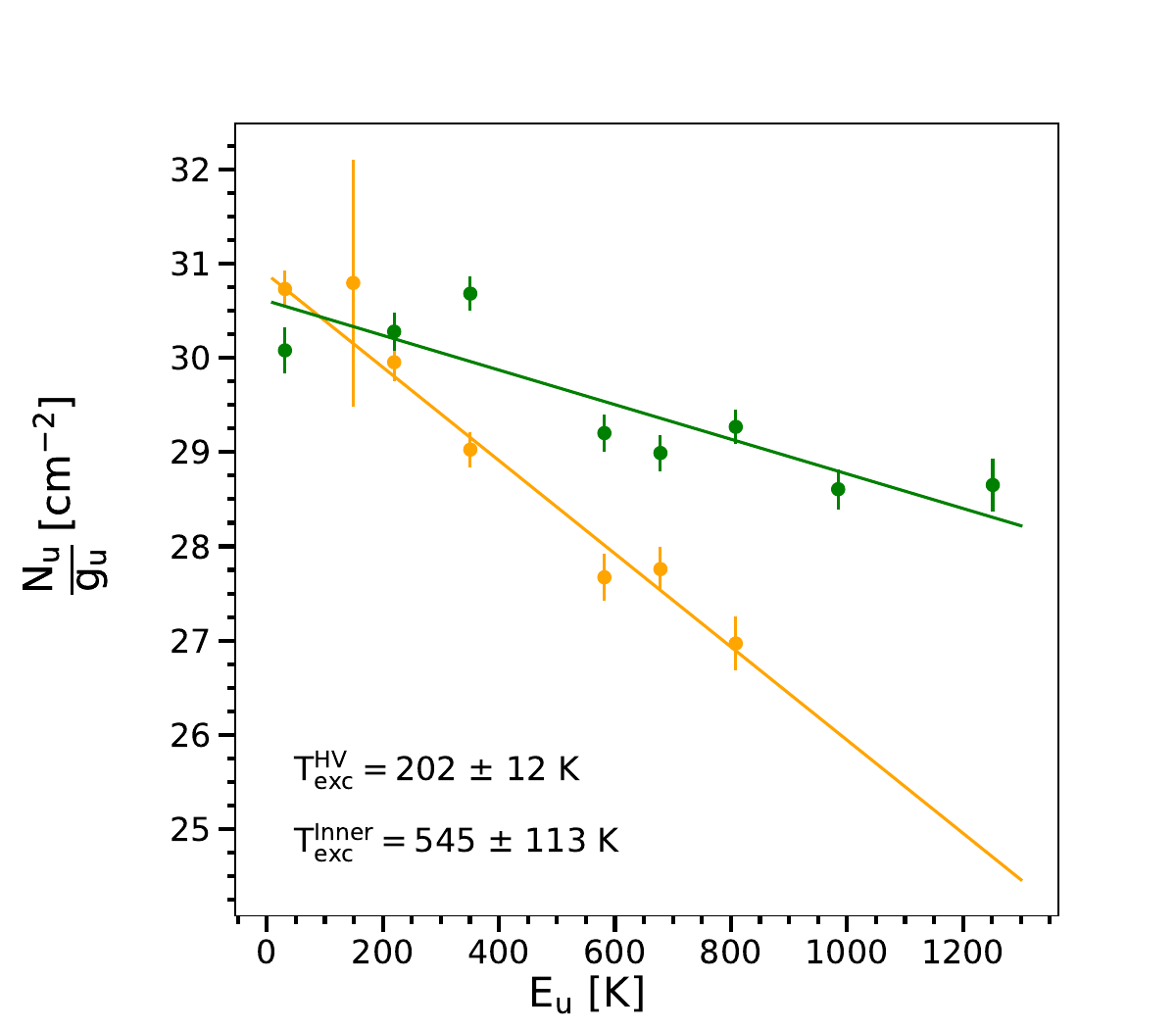} 
% \vspace*{-1.0 cm}
 \caption{Rotational diagram for SO$_2$ lines towards inner blob (green) and high-velocity blob (orange).}
   \label{fig:SO2}
\end{center}
\end{figure}

Similarly to the analysis of the inner northwest blob, we use seven SO$_2$ lines
in the ALMA Band 6 and Band 7 spectra towards the high-velocity blob to estimate the excitation temperature of SO$_2$.
We find an excitation temperature of
$202\pm12$~K (yellow points in Fig.~\ref{fig:SO2}).
Assuming the excitation temperature for CO to be the same of SO$_2$,
we derive a gas mass of $\sim 6 \times 10^{-8}$~M$_\odot$. 
Additionally, the assumed excitation temperature of $\sim 200$~K is equal to the gas kinetic temperature
in our model at $\sim16~R^{\star}_{887~\mu\rm{m}}$ from the star,
implying an angle between the blob position and the plane of the sky $\sim 45^\circ$.
For an angle ranging from 30$^\circ$ to 60$^\circ$,
we estimate the radial expansion velocity of this blob ($\upsilon_{\rm blob}$), its radial distance from R~Dor
($\Delta r_{\rm blob}$) and the time since its ejection ($\Delta t_{\rm blob}$) to be $25 < \Delta r_{\rm blob} < 43$~AU
($13.5< \Delta r_{\rm blob} < 23.4~R^{\star}_{887~\mu\rm{m}}$), $ 16 \lesssim \upsilon_{\rm blob} \lesssim 28$~km/s
and $4 \lesssim \Delta t_{\rm blob} \lesssim 12$~years.
Combined with our mass estimate,
this leads to mass-loss rates between $\sim 5 \times 10^{-9}$ and $\sim 2 \times 10^{-8}$~M$_\odot$/yr related to
the process that created the blob.
Assuming the blob to be roughly as thick in depth as it is in width, we estimate an average gas density of $\sim 4 \times 10^{7}$ H$_2$ molecules
per cm$^3$ in the blob.

\section{Discussion}
\label{sec:disc}

\subsection{Systemic velocity}
\label{sec:disc_vel}

The discrepancy between the systemic velocity obtained from fitting the CO~$v=0, J=2-1$ line observed using the ACA ($\sim 6.7$~km/s) and from
fitting the sharp absorption
features and the emission in the CO lines ($\sim 7.4$~km/s) is noteworthy, but
a systemic velocity of $\sim 7$~km/s is a reasonable compromise between these two values given the uncertainties.
Nonetheless, the position of the sharp absorption feature discussed in Section~\ref{sec:absorption}
and its mismatch with respect to the expected position (at $\upsilon = \upsilon_{\rm sys} - \upsilon_{\infty}$) indicates that
there might be an underlying cause for the difference in derived systemic velocity between the two methods.

One possibility is that the expansion velocity in the large-scale outflow adopted by us is too large. A maximum expansion velocity of 
4.8~km/s would lead to a consistent value of $ \upsilon_{\rm sys} = 6.7$~km/s both for the fit to the ACA line and the absorption feature.
However, this would produce a too-narrow CO~$v=0, J=2-1$ line in comparison to the profile obtained with the ACA.

Another possible explanation is an asymmetry in the velocity field of the large-scale outflow. If the approaching hemisphere
of the envelope has an average velocity towards us of 4.8~km/s (and, accordingly,
the receding hemisphere has a velocity of 6.2~km/s), the fit to the ACA line
and the absorption feature would both be consistent with a systemic velocity of $\sim 6.7$~km/s.
This scenario would not explain the improved fit
to the lines produced in the inner circumstellar envelope when
using the higher systemic velocity ($\sim 7.4$~km/s). Since the observed shift is relatively small
in velocity compared to the gas motions observed in the inner circumstellar envelope, this could most likely be overcome by modifying the velocity field
in the model. Nonetheless, we refrain from exploring this in this study.
The most straightforward way to assess whether there is an asymmetric velocity distribution
in the large-scale envelope is through a detailed analysis of the ACA maps. Such an investigation of the large-scale outflow is out of the scope of the present study.

\subsection{Stochastic velocity in the large-scale outflow}
\label{sec:disc_sto}

The value of the stochastic velocity we measure ($\lesssim~0.4$~km~s$^{-1}$)
can be compared to previous measurements of $0.7 \pm 0.1$ km~s$^{-1}$
in the outflow of W~Hya \citep{Khouri2014a}, and of 1.5~km/s \citep{DeBeck2012},
0.9~km/s \citep{Huggins1986} and 0.35~km/s \citep{Morris1985} in the outflow of CW~Leo.
Hence, we find a relatively low value comparable to that found by \cite{Morris1985} towards CW~Leo.
Our method of determination is very direct because it relies solely on the broadening of the absorption feature produced by gas in the large-scale
outflow.
The study of W~Hya relied on the shift of the line centres which is also affected by uncertainties on other parameters (such
as the assumed systemic velocity). The studies of CW~Leo mentioned above relied on the line shapes of the large-scale CO emission,
which is subject to uncertainties on asymmetries of the density distribution and/or velocity field.

Unfortunately, it is often unclear whether the reported values correspond to the standard deviation or the full-width at half maximum
of the gaussian line profiles. The upper limit of $\lesssim~0.4$~km~s$^{-1}$ for the standard deviation we report corresponds to
values of $\lesssim~0.95$~km~s$^{-1}$ for the full-width at half maximum. Hence, differences in the quantities
being reported might
account for at least part of the spread found in the literature.

\subsection{Circumstellar gas mass}

Our model shown in Fig.~\ref{fig:COmodel} implies a gas
mass of $\sim 3 \times 10^{-5}$~M$_\odot$ and $\sim 2 \times 10^{-6}$~M$_\odot$ in the inner and intermediate regions, respectively.
{ By integrating the profiles of the acceptable models from our sensitivity study (shown in Fig.~\ref{fig:grid}),
we obtain $\sim (3\pm1) \times 10^{-5}$~M$_\odot$ for the inner region
and $\sim 1.4 \times 10^{-6}$~M$_\odot$, with
an uncertainty of a factor of two, for the intermediate region. { Hence,
our best model provides gas masses in the inner and intermediate
regions which are very close to the average values obtained from the accepted models from the sensitivity study.}
These values are to be compared to the average mass-loss rate of R~Dor $\sim 10^{-7}$~M$_\odot$/yr.
The crossing times for a velocity of 5~km/s are $\sim 1$ and $\sim 6$~yrs for the inner and intermediate regions, respectively.
Therefore, for reasonable outflow velocities, these layers would sustain mass-loss rates of $> 10^{-5}$~M$_\odot$/yr (inner)
and $\sim 2 \times 10^{-7}$~M$_\odot$/yr (intermediate).
If the current gas density distribution is representative of the inner circumstellar environment of R~Dor, these values suggest that
the vast majority of gas in the innermost region would fall back to the star or remain in the extended atmosphere for hundreds of years,
while a significant fraction (or even all) of the mass in the intermediate region would be expelled. 
Therefore, our circumstellar model is consistent with the transition between the inner and intermediate regions (at $1.6~R^{\star}_{887~\mu\rm{m}}$)
corresponding to the transition between extended atmosphere and wind.

{ The (gravitationally bound) extended atmosphere is reminiscent of the gravitationally bound dust
shells invoked to explain aluminium oxide dust emission in W~Hya and R~Dor 
\citep{Khouri2015,VandeSande2018}, and also found in wind-driving models \citep{Hoefner2016}. 
The aluminium oxide dust mass inferred for
the gravitationally bound dust shell in R~Dor \citep[$\sim 6 \times 10^{-10}$~M$_\odot$,][]{KhouriPhD}
translates to a gas mass of $\sim 4 \times 10^{-6}$~M$_\odot$ assuming solar composition 
\citep{Asplund2021} and full aluminium condensation. Hence, we find
a lot more gas in the inner regions than 
required by the gravitationally bound dust shell models.
However, a conclusive correspondence of these two structures requires determining the
distributions of the aluminium 
oxide dust and the gas from spatially resolved observations carried out close in time.}

\subsection{Temperature profile}

The very steep temperature profile indicates very efficient cooling in the innermost layers of the circumstellar envelope of R~Dor.
The steepness ($\epsilon$) is constrained mostly from the ratio between absorption and emission as a function of aperture radius.
In the following paragraphs, we discuss potential alternative ways to reproduce the observations without resorting to a steep temperature profile,
 and make comparisons to other observations and
expectations from hydrodynamical models.

{ It is important to note that} our models for the $^{13}$CO line imply a slightly shallower
temperature profile in the inner region { than the CO,~$v=0$ line},
because more emission is seen with respect to absorption
than our model produces. Since the $^{13}$CO~$v=0, J=3-2$ line was observed about two weeks later than the CO~$v=0, J=2-1$ one, this could be at least partly due to variability.
In such a warmer model, with an exponent for the temperature in the inner region of 1.6 instead of 2.0, the fit to the $^{13}$CO~$v=0, J=3-2$ line is satisfactory when considering a  $^{12}$CO/$^{13}$CO of 10 { reported by
\citep{Ramstedt2014}, instead of the value of 15 required by our best model.}
This difference in the temperature profile corresponds to gas temperatures at $1.6~R^\star_{887\mu{\rm m}}$ higher by about 200~K ($\sim 20\%$).

{ Lines that are more optically thick in reality than in our models}
could increase the strength of absorption with respect to emission
without the need to make the gas cooler.
We note that the inner regions of our model reach indeed relatively high optical depths ($\lesssim 10$) in the CO~$v=0, J=2-1$ line.
However, two { facts speak against even higher optical depths being responsible for the
relatively strong absorption}.
First, the four transitions modelled by us have very different optical depths for
a given line of sight, because of differences in level populations of the given transition or abundance of the given molecule.
Second, an inspection of the cubes observed by ALMA reveals that in the CO~$v=0, J=2-1$ line (expected to reach the highest optical depths), the
maximum brightness temperature in the bright ring is $\sim 450$~K. For optically thick emission, the brightness temperature
of spatially resolved structures should reflect the actual source function, which in turn should be the same as the gas kinetic temperature
for the $J=2-1$ transition at such high gas densities.
Therefore, this measured brightness temperatures (which are well reproduced by our model) also imply relatively low gas
kinetic temperatures { close to the star even if the optical depths in our models are underestimated}.

Another alternative for the steep temperature profile is a significant asymmetry in the gas densities, with more gas present in
lines of sight towards the star than offset from it. Considering that we identify a few blobs in the vicinity of R Dor, this is not entirely
unlikely. Although we cannot rule out this scenario, we disfavour it at this point because it requires a specific configuration
of the envelope. Multi-epoch observations of R Dor and/or other targets can help determine whether density fluctuations
around the star significantly affect the ratio between observed absorption and emission.

These considerations suggest that the steep temperature profile in the innermost region
is indeed a real feature of the envelope at the time of the observations.
In fact, a comparison with other observational results indicates that steep temperature profiles might be a widespread feature of the circumstellar environments of oxygen-rich AGB stars.
Observations of the CO~$v=1,J=3-2$ line towards Mira \citep{Khouri2018} and W~Hya \citep{Vlemmings2017} also reveal low, $\lesssim 900$~K, 
excitation temperatures in the inner circumstellar environment ($\lesssim 2~R_\star$). Even if we find that the CO~$v=1$ levels are likely sub-thermally excited for the gas densities obtained in these studies,
the excitation temperature derived for them is the rotational-excitation temperature within the $v=1$ level, which should
be close to the kinetic temperature at these gas densities \citep{Vlemmings2017,Woitke1999}.

Spatially resolved observations of the brightness temperature of the continuum using ALMA also reveal a temperature profile for R~Dor
within $< 1.05 R^{\star}_{887~\mu\rm{m}}$
which is significantly steeper than that of a grey atmosphere model \citep{Vlemmings2019}. The continuum brightness temperature reported at
$\sim 1.035 R^{\star}_{887~\mu\rm{m}}$ of $\sim 2000$~K is even lower than we find at the same radial distance
($\sim 2240$~K). In fact, the difference between our temperature profile and a grey model
over this radial distance is small { (see Fig.~\ref{fig:model})}. We note that the observations of R~Dor reported by
\citep{Vlemmings2019} were acquired over one and a half months, and stellar variability might have played a role in the observed variations.

Comparisons to predictions from theoretical models is another way to assess whether these findings are expected.
However, such comparisons are not simple because the observations we report provide only one snapshot of the properties of the inner envelope.
Hence, we investigated whether dynamical models
can at all produce temperature profiles as steep as that inferred by us. Our results imply that the
kinetic temperature decreases by a factor $\sim 2.5$ between
$R^{\star}_{887~\mu\rm{m}}$ and 1.6~$R^{\star}_{887~\mu\rm{m}}$, with the temperature at $R^{\star}_{887~\mu\rm{m}}$ being 2370~K.
The most extreme temperature profiles presented by \cite{Ireland2011} decrease by at most a factor of 2.2 between the radius at which
 they reach 2370~K
and a 60\% larger radius. The profiles shown by \cite{Freytag2023} and \cite{Hoefner2016} have shallower profiles with factors of at most $\sim 1.7$
over the relevant range. This comparison indicates that existing dynamical models might be able to reproduce the very steep profile we observe in certain phases
of specific models, but not often. If future observations indicate that such steep profiles are the rule,
more efficient cooling in dynamical models would be necessary.
This would likely have a significant impact on the calculations of dust growth and wind driving in such models.

\subsection{Blobs}

\subsubsection{Inner blob}

The properties of the inner blob identified in the maps seem consistent with bubbles created from convection in dynamical models \citep[e.g.][]{Freytag2023}, but the inferred expansion
$\sim 13$~km/s velocities seem rather high at a fairly large distance from the star ($\sim 4.6~R^{\star}_{887~\mu\rm{m}}$) in comparison with
the observed terminal velocity of the outflow, $\sim 5.5$~km/s. A study of the dust content of this inner blob in comparison to the gas properties derived here will be
performed in an upcoming publication (Schirmer et al., {\it in prep.}).
The gas mass of this blob implies that a similar blob needs to be ejected at most every few years to account for the mass-loss rate observed on larger scales.

\subsubsection{High-velocity blob}

The high-velocity blob seen to the southeast in the CO~$v=0, J=2-1$ line stands out as an unexpected feature of the circumstellar envelope of R~Dor. We assume the
blob to be moving radially away from the star, because it seems unlikely that it would have a high tangential velocity (as argued in Section~\ref{sec:highV_blob}).
The density estimated from the CO~$v=0, J=2-1$
line flux is at least one order of magnitude larger than the densities expected based on our model (not including blobs).
 
 Although the SO$_2$ excitation temperature in the blob is quite well constrained by our population diagram
 ($202\pm12$~K) there is considerable uncertainty on the blob
 position because the kinetic temperature is expected to vary slowly with radius at these distances from the star. Also, { for both blobs}
 the excitation temperature of SO$_2$ might differ
 from the kinetic temperature, and the kinetic temperature within the blob might differ from that of the surrounding gas (and, hence, from our model) given
 the difference in density.
For reference, the kinetic temperature in the model varies from 220~K to 160~K between 13.5~$R^{\star}_{887~\mu\rm{m}}$ and 23.4~$R^{\star}_{887~\mu\rm{m}}$.
The most direct way to constrain the position
of the blob would be new measurements to determine its proper motion, given the short estimated ejection times between 4 and 12~years.

\subsection{Velocity distribution and wind-driving mechanism}

Our results reveal a complex velocity distribution around the star, with outflow and infall of gas in the innermost region. Moreover, we find departures from spherical symmetry both
between the near and far hemispheres of the innermost layer, as well as in the absorption profile observed against the star. The fact that we observe different absorption profiles
against the stellar disc, which is resolved by only a few resolution elements, indicates that there is a relatively large-scale component of the velocity distribution with scales comparable to that of the star. The conclusion
that the innermost region is dominated by infall in the near hemisphere and outflow in the far one points in the same direction. { We attribute the gas velocity distribution in the inner region to be mostly a
consequence of convective motions and pulsation of the star.
Although a much faster rotation than expected for an isolated
AGB stars has been reported for R~Dor \citep{Vlemmings2018}, the projected velocities along the line
of sight are much smaller than the radial velocities we infer, and should not affect significantly the inflow
and outflow velocities we report. We find no evidence in the data examined by us of a disk around R~Dor,
as suggested by \citep{Homan2018}.}

Outside the inner, dense region, the circumstellar envelope seems to transition quickly to the outflow which is observed on large scales. From the absorption profile,
we find a relatively fast layer of outward-moving gas, which might be a shock front as expected from dynamical models \citep[e.g.][]{Hoefner2019}. As discussed above, it is not straightforward to
determine the exact location of the gas producing this absorption, but the fact that it is not seen in the $v=1$ lines indicates that it is located at $r > 1.6~R^{\star}_{887~\mu\rm{m}}$. Regarding the acceleration profile,
we assume that the terminal velocity of the large-scale outflow is reached only at 10~$R^{\star}_{887~\mu\rm{m}}$ in our model, but the data are most likely also consistent with faster acceleration. However, this was not explored in our calculations.

We introduced the two modifications to the velocity field only on the far side of the envelope so that they do not affect the absorption profile produced by our model, which was the starting point of the modelling procedure
and fits the absorption profiles very well.
It is not surprising that the velocity field of the model we derived based on the absorption observed against the star does not fit the velocity field in the 
far hemisphere. However, it is somewhat surprising that the two hemispheres would have, on average, the opposite behaviour,
as shown in Fig.~\ref{fig:model}. While in the near side of the envelope the inner and intermediate regions in our model are falling towards and moving away from the star, respectively, in the far side of envelope they flow outwards and fall back, respectively, instead. As discussed in Section~\ref{sec:absorption}, we see signatures of different velocity components in the absorption profile of the CO~$v=1, J=3-2$ line,
which indicates that the far-side vs near-side division is only a rough approximation. Hence, the exact relative size and distribution of the
infalling and outflowing components are not strongly constrained. 
Nonetheless, the scenario in which the far and near sides of the envelope have behaviours which on average are quite different is a good description of the velocity field that reproduces the line profiles well. 

The importance of the gas blobs to the wind-driving process is not clear from the analysis we have carried out.
While the inner blob could be produced by convection, the process that produced the high-velocity blob is less clear.
It could be driven by radiation
pressure on dust, in principle, because wind-driving models reach comparable outflow velocities of $\sim 20$~km/s \citep[e.g.][]{Bladh2019}.
However, it is unclear why this one blob would have velocities so much higher than those
of the large-scale outflow ($\sim 5.5$~km/s) at such large distances, or how the interaction between this blob and the slower moving gas proceeds.
\cite{Decin2018} argued that a high-velocity component exists in the circumstellar envelope of R~Dor, either because of blobs or acceleration of the large-scale outflow.
An inspection of the central-beam ACA spectrum we use for constraining the large-scale outflow { (Fig.~\ref{fig:CO2-1_ACA})}
reveals weak wings extending to maximum expansion velocities $\sim - 17$~km/s on
the blue-shifted side of the CO~$v=0, J=2-1$ line with flux densities $< 100$~mJy for velocities $< -9$~km/s. The CO~$v=0, J=2-1$
line extracted from an aperture with 240~mas diameter reveals flux densities at $\pm 10$~km/s of $\sim 50$~mJy, while the spectrum of the high-velocity blob peaks at $\sim 50$~mJy
at $\sim -14$~km/s. Hence, the combination of emission from the extended atmosphere and the high-velocity blob seem to account for all observed high-velocity emission detected
in the ACA spectrum. If a significant larger mass of high-velocity gas existed, stronger emission should have been picked up by the ACA observations.

Based on these considerations, we conclude that the
high-velocity blob observed is either a very rare event that will propagate through the envelope at its relatively high velocity, or that such ejections
happen on timescales more comparable to the inferred ejection timescale ($\sim 10$~years) and that the produced blobs get slowed down on a comparable timescale. Although we
cannot reach a definitive conclusion, the second scenario seems more plausible both based on the unlikelihood of observing a rare event and on the
hydrodynamical interactions between blob and outflow, { which must act to decelerate the blob on long timescales}.
If this is indeed the case, this blob should be observed to dissipate in the near future and similar blobs should be created.
We point out that a more complex morphology is seen in the $^{29}$SiO $v=0, J=5 - 4$ line \citep{Vlemmings2018},
which could imply this to be a feature that is stable on somewhat longer timescales than expected in the scenario explored by us.

The inner blob reported by us is a good candidate for
producing a new high-velocity blob. Therefore, future observations of R~Dor will shed light on the nature of these features,
and will allow the study of the processes controlling the wind driving in real time.

\section{Summary}
\label{sec:summary}

We report observations using ALMA of the AGB star R~Doradus. The high-angular resolution of the observations allows even for the stellar
photosphere at sub-mm wavelengths to be spatially resolved with uniform disc size of $\sim 62$~mas, corresponding to
$\sim 1.835$~AU at 59 pc.

We investigate the properties of the gas in the extended atmosphere around the continuum source using lines
$v=0, J=2-1$ and the $v=1, J=2-1$ and $3-2$ lines of $^{12}$CO, and $v=0, J=3-2$ of  $^{13}$CO.
Based on the absorption produced by the cooler gas against the star,
we determine that gas in the near-side hemisphere can be divided in two layers with different average velocities. The innermost layer is falling back
to the star,
while the layer above that is moving outwards.
The Band~7 observations of the CO~$v=1, J=3-2$ and $^{13}$CO~$v=0, J=3-2$ lines, which were acquired at higher spatial resolution, also show how both
the intensity of absorption and the velocity distribution of the absorbing gas change across the stellar disc. The absorption profile of the CO~$v=0, J=2-1$ line also shows a sharp feature
which we attribute to absorption produced in the large-scale outflow. Based on the profile of this feature, we constrain the standard deviation of the stochastic velocity distribution in 
the outer wind to be $\lesssim 0.4$~km/s.

Based on the velocity field inferred from the absorption profiles, we construct a model to fit the observed lines using the 3D molecular excitation and line radiative transfer code LIME.
We find temperature and density profiles that are very steep close to the star and become shallower at radii $> 1.6~R^{\star}_{887~\mu\rm{m}}$.
While the innermost region contains a lot of mass $> 10^{-5}$~M$_\odot$, most of which will fall back to the star or remain in this region for centuries before becoming gravitationally unbound,
the circumstellar envelope seems to quickly transition to gas densities consistent with those expected from the mass-loss rate reported
in the literature.
We find that our symmetric model including rotation fails to reproduce some aspects of the observed
line positions and shapes. This is solved by reversing the velocity field between the innermost and intermediate layers in the far side of the envelope.
A disagreement between the systemic velocity determined based on the large scale CO~$v=0, J=2-1$ line profile and based on the profile of the absorption against the star
leads us to speculate that the velocity field of the large-scale envelope might be asymmetric.

We characterise two blobs found at small angular separations from the central star. One of them,
seen very close to the central star (at a distance of $\sim 4.5~R^{\star}_{887~\mu\rm{m}}$),
resembles structures created in hydrodynamical models through the action of convection, but we are unable to determine its formation mechanism.
We find an excess mass in this blob with respect to its surroundings of $\sim 6 \times 10^{-8}$~M$_\odot$, which implies that blobs such as this
have to be created every few
years to have a substantial effect on the mass-loss rate of R~Dor. The second blob,
most likely located at a radial distance
of $\sim 16~R^{\star}_{887~\mu\rm{m}}$, is moving
at a considerably higher velocity ($\sim 20$~km/s) than expected at these distances from the star. Surprisingly, we
also infer it to contain a gas mass of
$\sim 6\times 10^{-8}$~M$_\odot$. We find it was created roughly during the last 10~years.

Understanding the properties of these blobs seems important for advancing our understanding of the wind-driving mechanism in AGB stars. Even if the mass-loss rate
associated with them turns out to be much lower than that derived for R~Dor, understanding the
different processes at play would undoubtedly provide significant insight into the relevant processes at play.
The fate of these blobs when moving through the slower-moving circumstellar envelope is at the moment unclear.
Monitoring their evolution would provide important new constraints.

\begin{acknowledgements}
{ The authors would like to thank the anonymous referee for the careful reading of the paper
and the constructive comments which helped improve the quality of this manuscript.
This paper makes use of the following ALMA data: ADS/JAO.ALMA\#2017.1.00191.S, ADS/JAO.ALMA\#2017.1.00824.S, and ADS/JAO.ALMA\#2017.1.00595.S.
ALMA is a partnership of ESO (representing its member states), NSF (USA) and NINS (Japan), together with NRC (Canada), MOST and ASIAA (Taiwan), and KASI
(Republic of Korea), in cooperation with the Republic of Chile. The Joint ALMA Observatory is operated by ESO, AUI/NRAO and NAOJ.
TK and VW acknowledge support from the Swedish Research Council through grants 2019-03777 and 2020-04044, respectively.
TS acknowledges support from the Knut and Alice Wallenberg Foundation through grant nr. KAW 2020.0081.
MS acknowledges support from the research council of Norway, project number: 335497.}
\end{acknowledgements}
\bibliographystyle{aa}
\bibliography{bibliography_2}

\begin{appendix}

\section{Model selection for grid calculation}
\label{app:grid}

The procedure of selecting models as acceptable fits consisted of ranking the models based on the value of the sum of the squared of the residuals in comparison to the best model from Section~~\ref{sec:firstModel}.
Then, the line profiles were inspected by eye.

The models shown in Fig.~\ref{fig:grid_good} and \ref{fig:grid_bad} exemplify why
the selection based on the residuals was not enough to select acceptable models.
For instance, while a model that produces too strong absorption for small apertures and too strong emission for larger ones in the blue-shifted wing might have a value of the squared residuals
that is comparable to a model that produces slightly too strong emission overall, the former model might be a clear worse fit for the region producing the emission and absorption of
the spatial region in question. In Fig~\ref{fig:layers}, we show models calculated with the parameters of the best model for the inner, intermediate, and outflow regions separately.
These can be used to train the eye to interpret how emission from the different regions contribute to the line profiles.
Of course, the addition of these profiles is not the same as the profiles of the given model because of optical depth effects.

\begin{figure*}[t]
% \vspace*{-2.0 cm}
\begin{center}
 \includegraphics[width=0.9\textwidth]{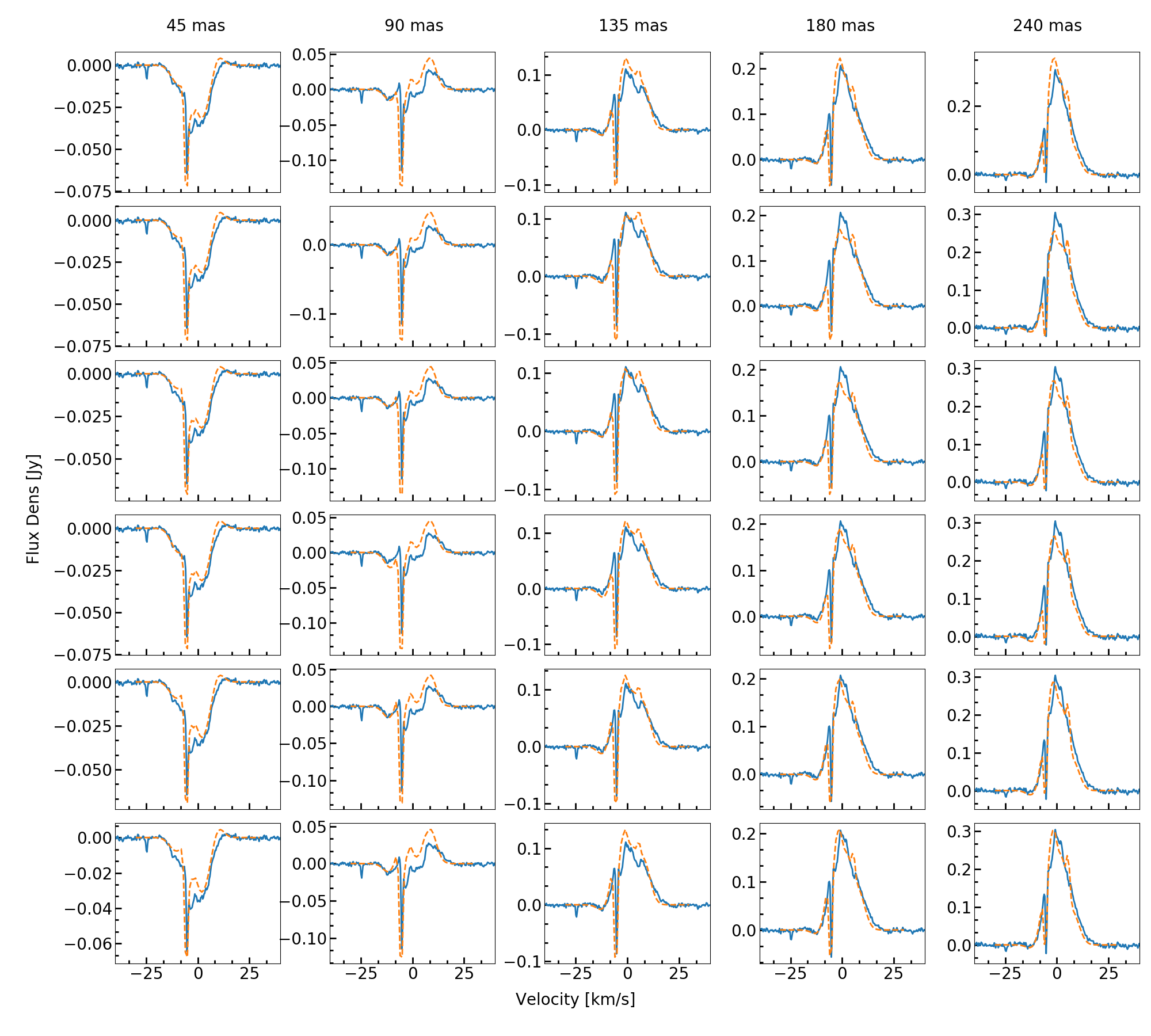} 
% \vspace*{-1.0 cm}
 \caption{Top-ranked models from our residual minimisation and considered good fits upon inspection of the line profiles.
 Observed spectra of the CO~$v=0, J=2-1$ line extracted from five different aperture sizes is shown by the blue histogram. From left to right, the apertures diameters
 are 45~mas, 90~mas, 135~mas, 180~mas, and 240~mas. The dashed yellow lines represent the models.
 The first row shows our best model from Section~~\ref{sec:firstModel}, while models from the grid considered acceptable are shown in
 the following rows.}
   \label{fig:grid_good}
\end{center}
\end{figure*}

\begin{figure*}[t]
% \vspace*{-2.0 cm}
\begin{center}
 \includegraphics[width=0.9\textwidth]{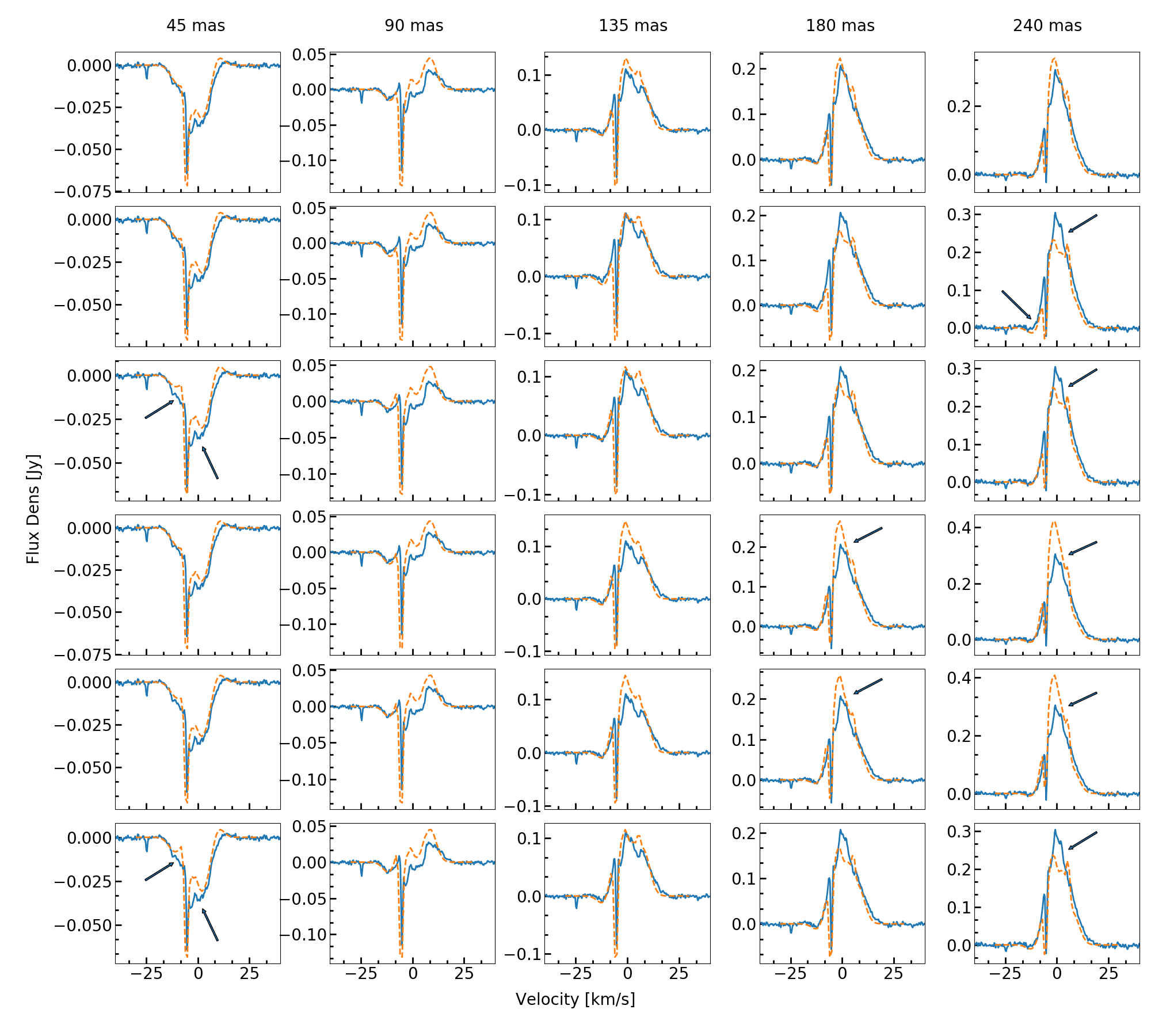} 
% \vspace*{-1.0 cm}
 \caption{Top-ranked models from our residual minimisation but considered not good fits upon inspection of the line profiles.
 Observed spectra of the CO~$v=0, J=2-1$ line extracted from five different aperture sizes is shown by the blue histogram. From left to right, the apertures diameters
 are 45~mas, 90~mas, 135~mas, 180~mas, and 240~mas. The dashed yellow lines represent the models.
 The first row shows our best model from Section~~\ref{sec:firstModel}, while models from the grid considered acceptable are shown in
 the following rows. The arrows indicate
 the main discrepancies between the given grid models and our best model that led to the classification as unacceptable fits.}
   \label{fig:grid_bad}
\end{center}
\end{figure*}
 
 \begin{figure*}
% \vspace*{-2.0 cm}
\begin{center}
 \includegraphics[width=\textwidth]{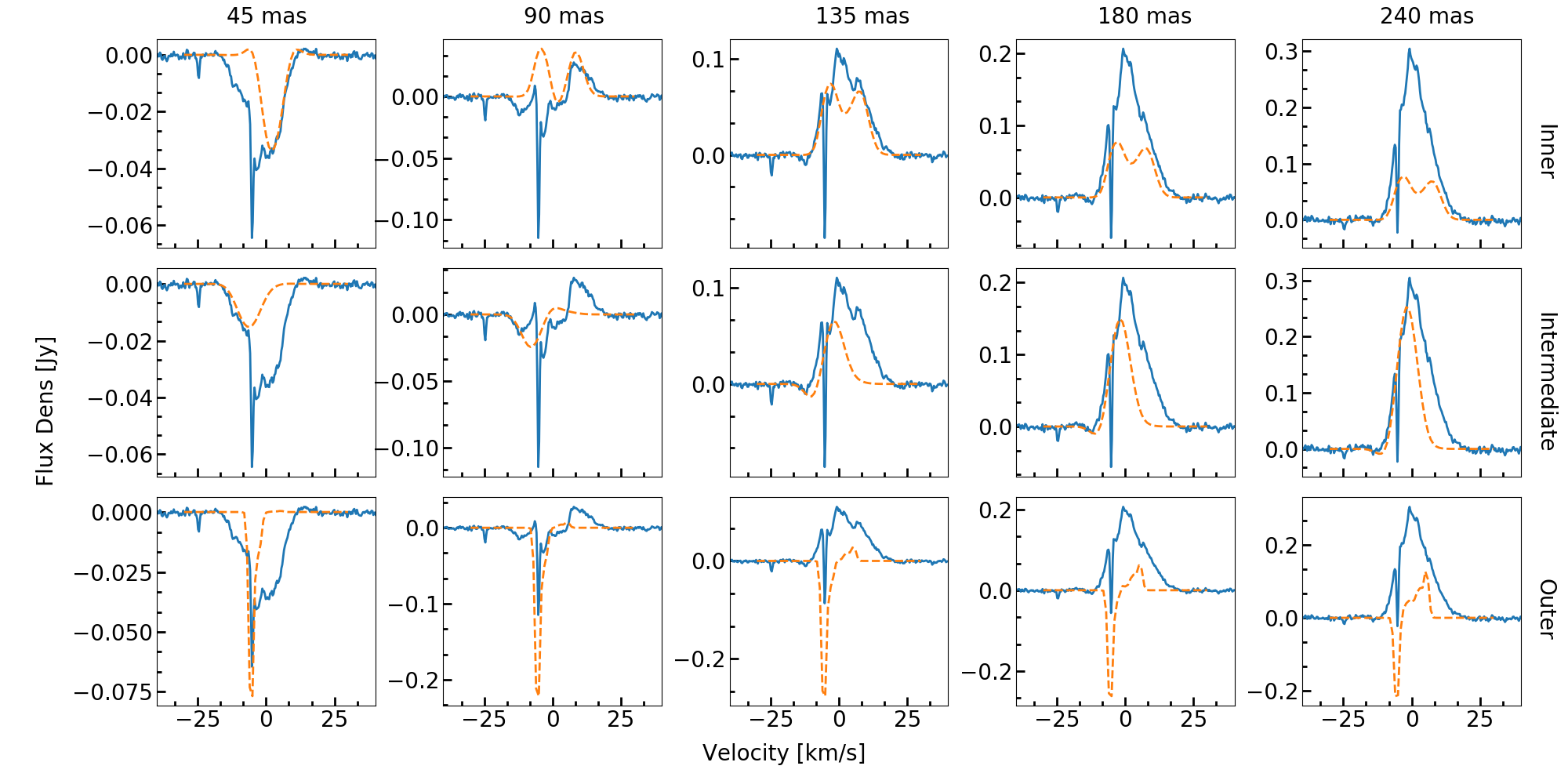} 
% \vspace*{-1.0 cm}
 \caption{Spectra of the CO~$v=0, J=2-1$ line extracted from five different aperture sizes is shown by the blue histogram. From left to right, the apertures diameters
 are 45~mas, 90~mas, 135~mas, 180~mas, and 240~mas. The dashed yellow lines represent the models which were calculated considering our best models and including only the inner layer (top), the intermediated layer (middle) and the outflow (bottom).
See text for details on the models.}
   \label{fig:layers}
\end{center}
\end{figure*}

\FloatBarrier

\section{Channel maps}
\label{app:channel_maps}

 \begin{figure*}[t]
   \centering
      \includegraphics[width=\textwidth]{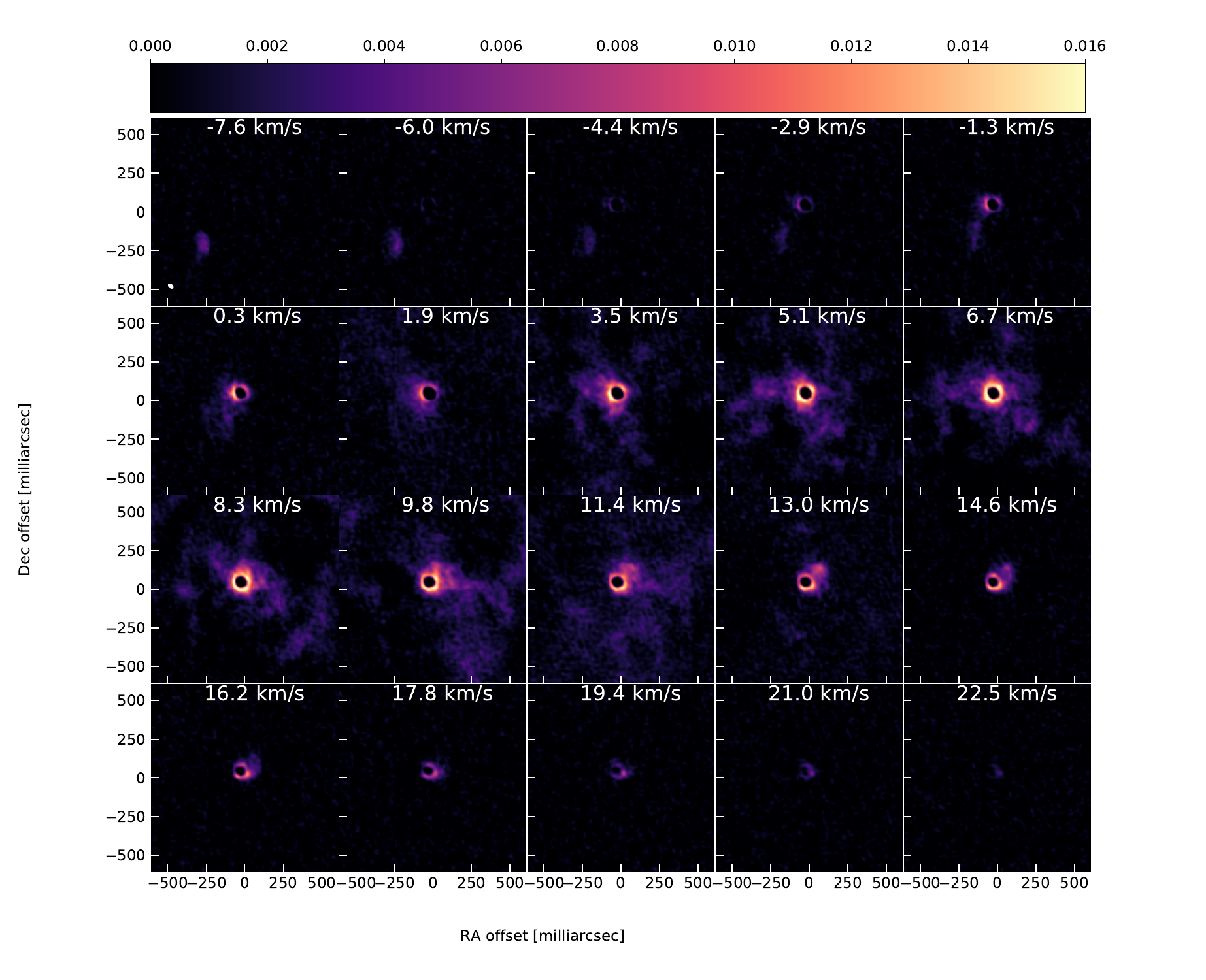}
      \caption{Observed channel maps of the CO~$v=0, J=2-1$ transition given in Jy/beam as a function of velocity.}
         \label{fig:COv0_channels}
   \end{figure*}

 \begin{figure*}[t]
   \centering
      \includegraphics[width=\textwidth]{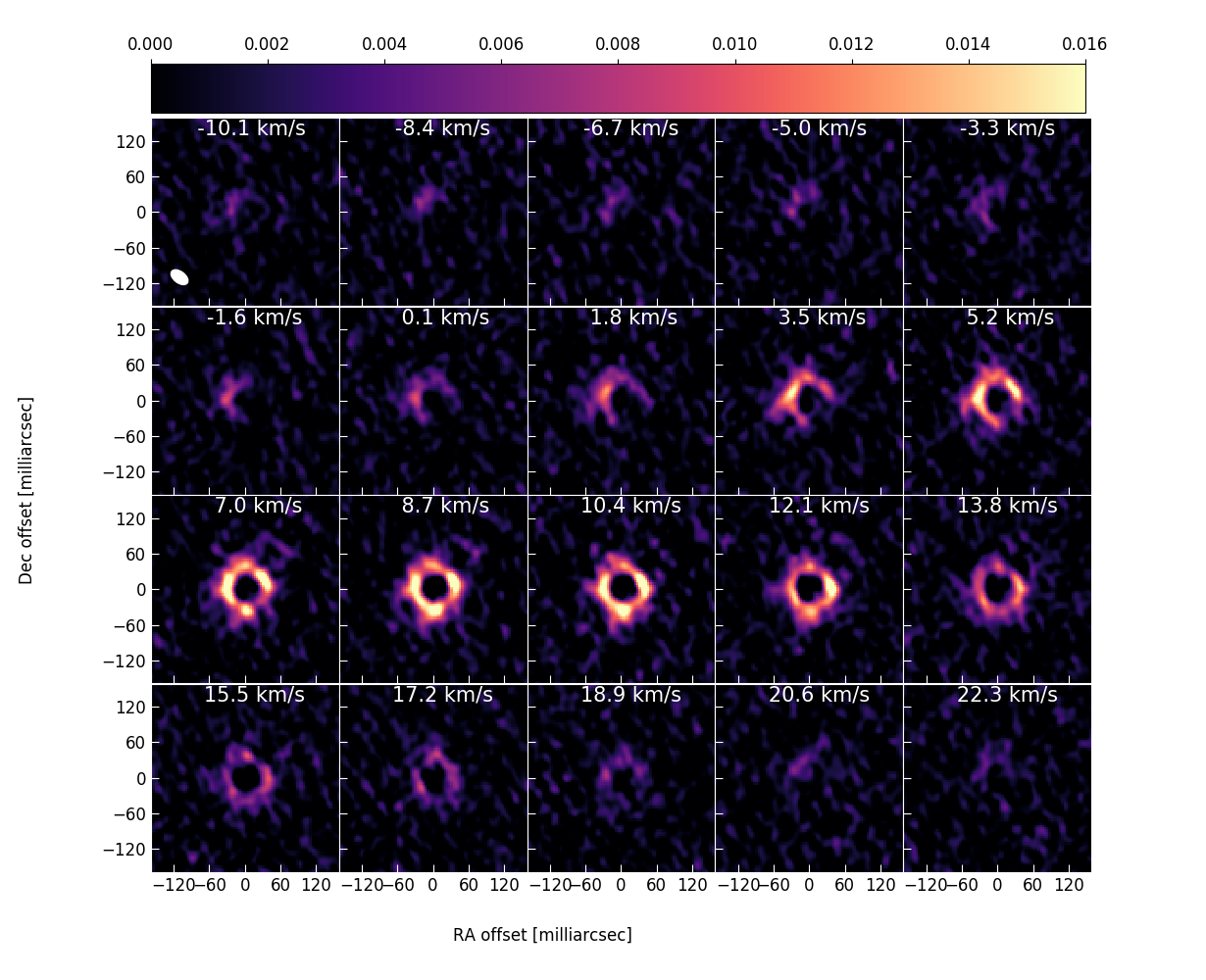}
      \caption{Observed channel maps of the CO~$v=1, J=3-2$ transition given in Jy/beam as a function of velocity.}
         \label{fig:COv1_channels}
   \end{figure*}

 \begin{figure*}[t]
   \centering
      \includegraphics[width= 18cm]{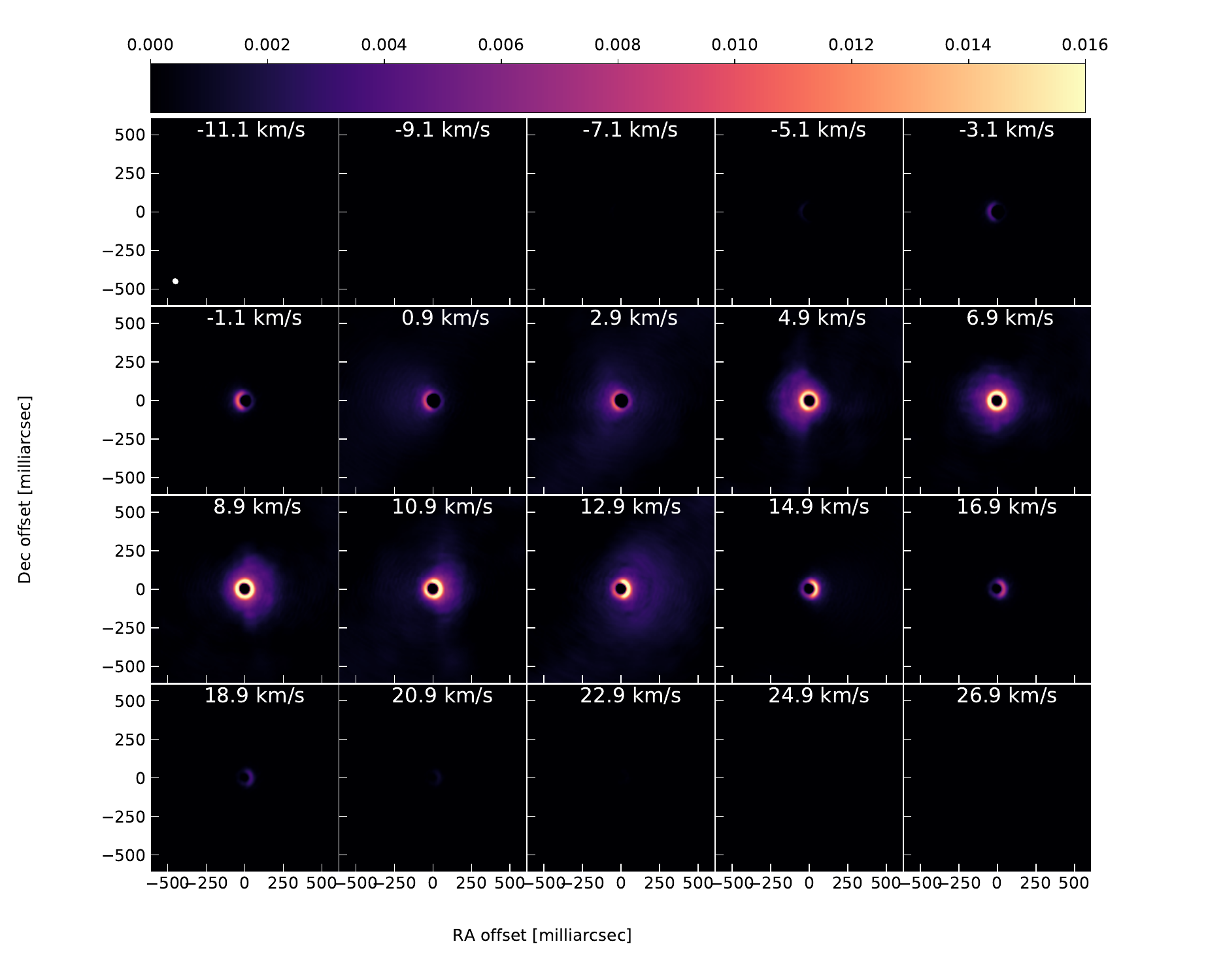}
      \caption{Channel maps for the CO~$v=0, J=2-1$ line in the best model with symmetric velocity distribution and rotation. The emission is shown in Jy/beam as a function of velocity.
      The white ellipse on the lower left corner of the upper right panel shows the size of the ALMA gaussian beam at half power. The models were shifted to a systemic velocity of 6.7~km/s.
      }
         \label{fig:modelCOv0_symm}
   \end{figure*}

 \begin{figure*}[t]
   \centering
      \includegraphics[width= 18cm]{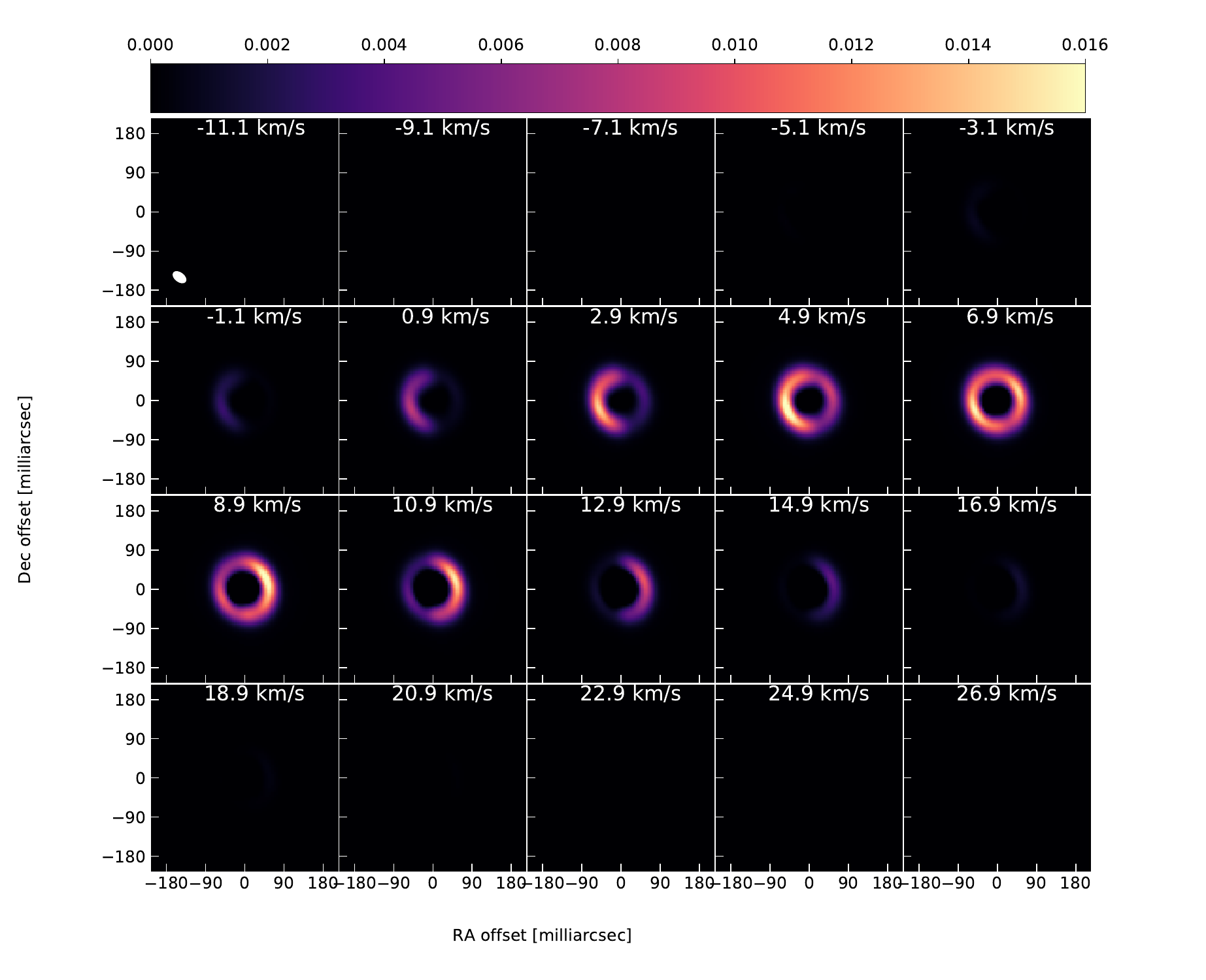}
      \caption{Channel maps for the CO~$v=1, J=3-2$ line in the best model with symmetric velocity distribution and rotation. The emission is shown in Jy/beam as a function of velocity.
      The white ellipse on the lower left corner of the upper right panel shows the size of the ALMA gaussian beam at half power. The models were shifted to a systemic velocity of 6.7~km/s.
      }
         \label{fig:modelCOv1_symm}
   \end{figure*}

 \begin{figure*}[t]
   \centering
      \includegraphics[width= 18cm]{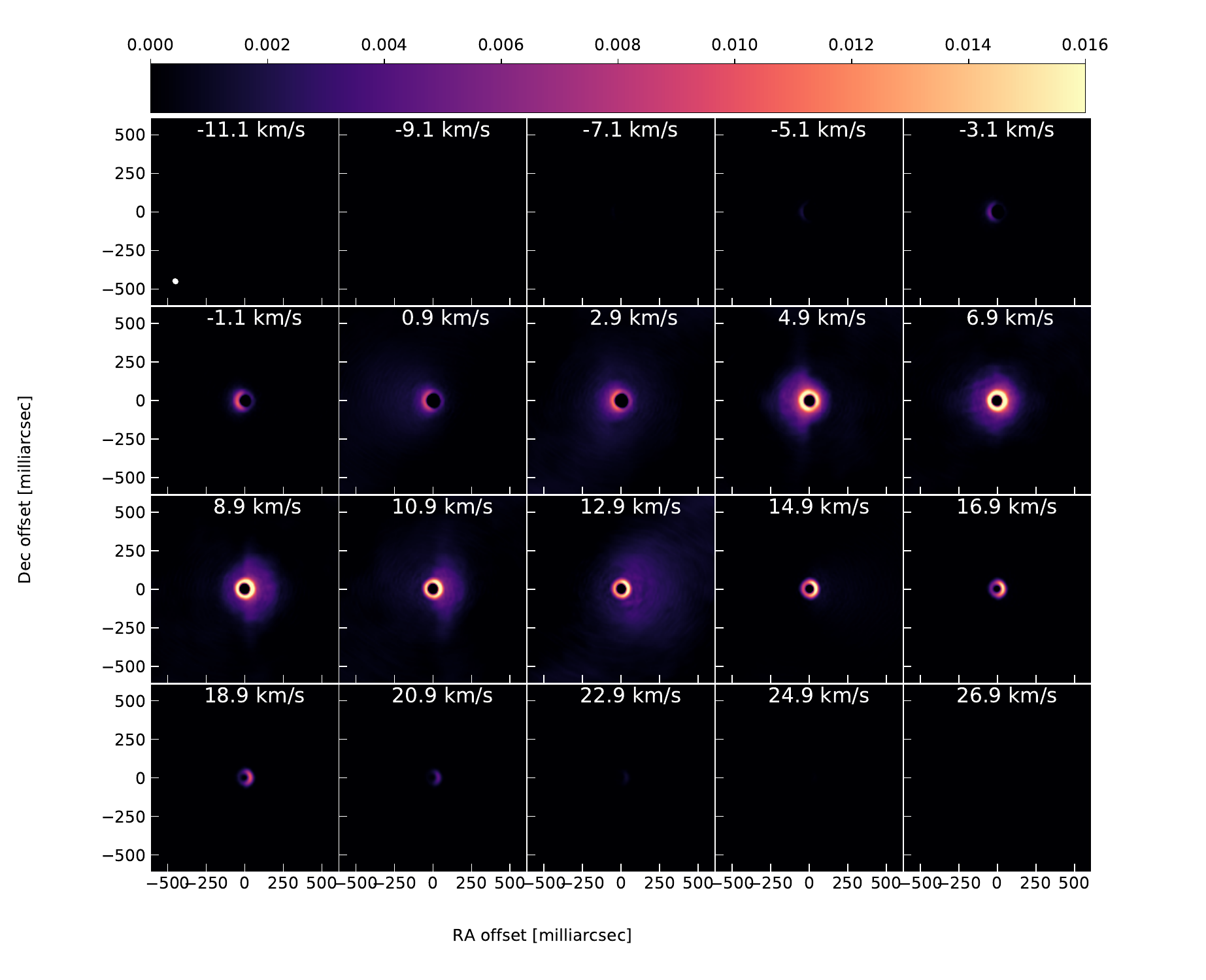}
      \caption{Channel maps for the CO~$v=0, J=2-1$ line in the best model with non-symmetric velocity distribution and rotation. The emission is shown in Jy/beam as a function of velocity.
      The white ellipse on the lower left corner of the upper right panel shows the size of the ALMA gaussian beam at half power. The models were shifted to a systemic velocity of 6.7~km/s.
      }
         \label{fig:modelCOv0_nonSymm}
   \end{figure*}

 \begin{figure*}[t]
   \centering
      \includegraphics[width= 18cm]{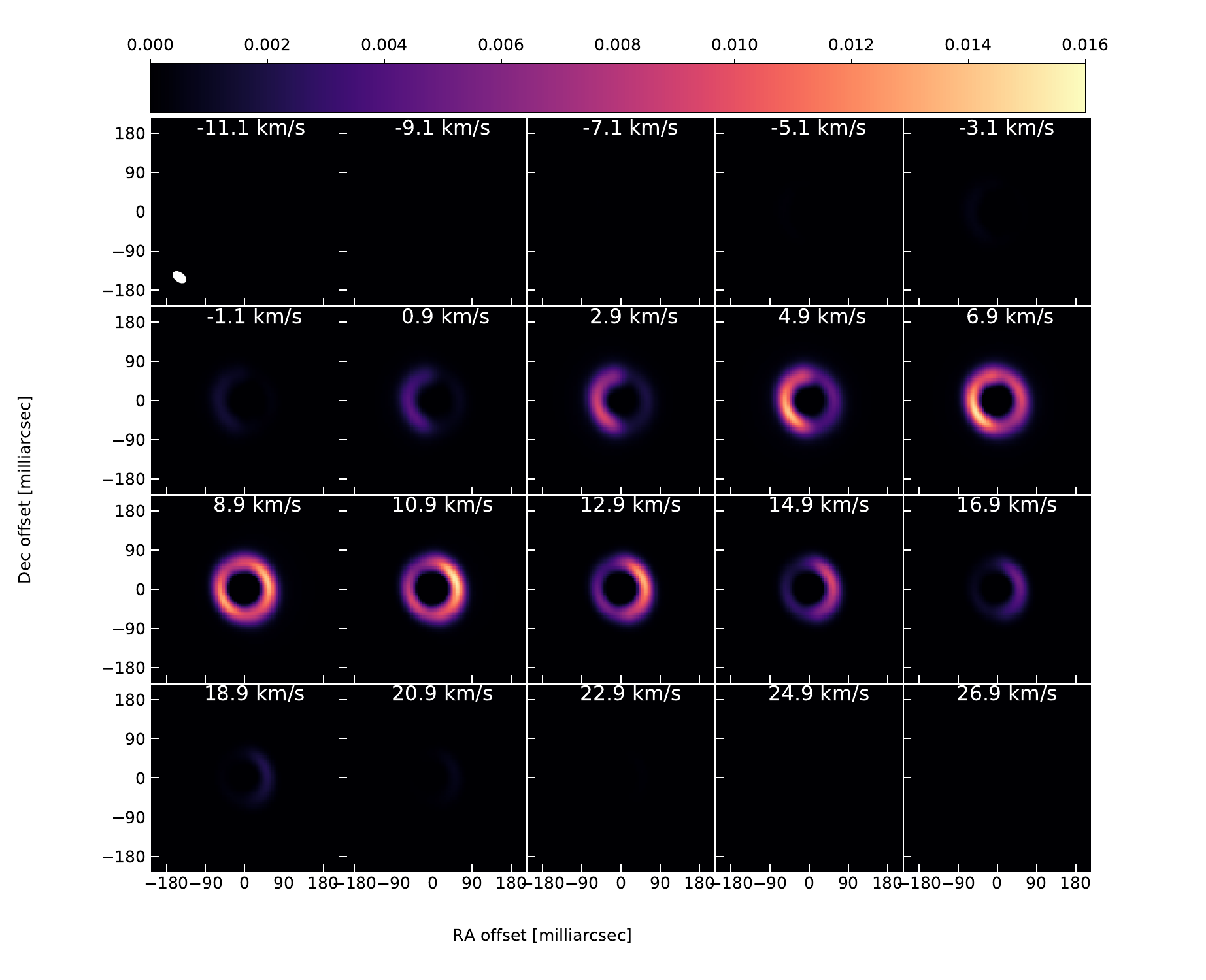}
      \caption{Channel maps for the CO~$v=1, J=3-2$ line in the best model with non-symmetric velocity distribution and rotation. The emission is shown in Jy/beam as a function of velocity.
      The white ellipse on the lower left corner of the upper right panel shows the size of the ALMA gaussian beam at half power. The models were shifted to a systemic velocity of 6.7~km/s.
      }
         \label{fig:modelCOv1_nonSymm}
   \end{figure*}

 \end{appendix}
\end{document}